\documentclass[conference,onecolumn,12pt]{IEEEtran}
\usepackage{verbatim,epsfig,times,subfigure}
\usepackage{amsmath,amsfonts,amssymb,mathrsfs}
\pdfoutput=1
\usepackage[pdftex]{color}
\usepackage{url}
\usepackage{hyperref}
\definecolor{brown}{cmyk}{0,0.81,1,0.60}
\definecolor{magenta}{rgb}{0.4,0.7,0}
\definecolor{gray}{rgb}{0.5,0.5,0.5}
\definecolor{red}{rgb}{1,0,0}
\definecolor{green}{rgb}{0.5,0,0.5}
\definecolor{blue}{rgb}{0,0,1}

\newcommand{\tbltxt}[1]{\parbox[t]{5.1in}{#1}}
\newtheorem{definition}{{\bf \hspace{-0.18in} Definition}}
\newtheorem{assumption}{{\bf \hspace{-0.18in} Assumption}}
\newtheorem{theorem}{{\bf \hspace{-0.18in} Theorem}}
\newtheorem{lemma}{{\bf \hspace{-0.18in} Lemma}}

\newtheorem{corollary}{{\bf \hspace{-0.18in} Corollary}}
\def\done{\hspace*{\fill} \rule{1.8mm}{2.5mm}}

\begin{document}

\title{On the Evolution of the Internet Economic Ecosystem }

\author{\IEEEauthorblockN{Richard T. B. Ma}
\IEEEauthorblockA{Department of Computer Science\\
National University of Singapore\\
Email: tbma@comp.nus.edu.sg}
\and
\IEEEauthorblockN{John C. S. Lui}
\IEEEauthorblockA{Department of CS \& E\\
Chinese University of Hong Kong\\
Email: cslui@cse.cuhk.edu.hk}\and
\IEEEauthorblockN{Vishal Misra}
\IEEEauthorblockA{Department of Computer Science\\
Columbia University, USA\\
Email: misra@cs.columbia.edu}
}



\maketitle

\begin{abstract}
The evolution of the Internet has manifested itself in many ways: the traffic characteristics, the interconnection topologies and
the business relationships among the autonomous components. It is important to understand why (and how) this evolution came about, and how the interplay of these dynamics may affect future evolution and services.
We propose a network aware, macroscopic model that captures the characteristics and interactions of the application and network providers, and show how it leads to a market equilibrium of the ecosystem.
By analyzing the driving forces and the dynamics of the market equilibrium, we obtain some fundamental understandings of the cause and effect of the Internet evolution, which explain why some historical and recent evolutions have happened.
Furthermore, by projecting the likely future evolutions, 
our model can help application and network providers to make informed business decisions so as to succeed in this competitive ecosystem.
\end{abstract}

\section{Introduction}\label{section:intro}
The Internet has been and is still changing unexpectedly in many aspects.
Started with elastic traffic and applications, e.g., emails and webpage downloading, we have seen significant rise in inelastic traffic, e.g., video and interactive web traffic, across the Internet.
According to \cite{craig10internet}, from $2007$ to $2009$, web content traffic had increased from $41.68\%$ to $52\%$, reaching more than half of the total Internet traffic.  From a network perspective, the Internet originated from government-owned backbone networks, i.e., the ARPANET, and then evolved to a network of commercial Autonomous Systems (ASes) and Internet Service Providers (ISPs). Meanwhile, ISPs formed a hierarchical structure and were classified by tiers, with higher tier ISPs cover larger geographic regions and provide transit service for smaller/lower tier ISPs. However, recent study \cite{flattening08} has reported that large content providers, e.g., Google and Microsoft, are deploying their own wide-area networks so as to bring content closer to users and bypassing Tier-1 ISPs on many paths. This is known as the {\em flattening phenomenon} of the Internet topology.

Changes in the content or network topology do not happen independently. Rather, they are driven by the changes in the business relationships among the players in the Internet ecosystem. Not surprisingly, we have observed dramatic changes in the
business relationships 
between the content providers and the ISPs and among the ISPs themselves.
Traditionally, ISP settlements were often done bilaterally under either a (zero-dollar) peering or in the form of a customer-provider relationship. Tier-1 ISPs, e.g., Level 3 \cite{L3}, often charge lower tier ISPs for transit services and connect with each other under settlement-free peering. However, the Tier-1 ISPs do not have any guarantee in their profitability as the Internet evolves. For instance, we have seen exponential decrease (around $20\%$ a year) in IP transit prices \cite{Norton-playbook}. Also, peering disputes happened, e.g., the de-peering between Cogent \cite{Cogent} and Level 3 in 2005, where the lower tier ISPs that are closer to content or users refused to pay for the transit charge.
This leads to the recent debate of network neutrality \cite{tim05nn}, which reflects the ISPs' willingness to provide value-added and
differentiated services and potentially charge content providers based on different levels of service quality.

The situation is further complicated by the emergence of new players in the ecosystem:  Content Delivery Networks (CDNs), e.g., Akamai \cite{Akamai} and Limelight \cite{Limelight}, and high-quality video streaming providers, e.g., Netflix \cite{Netflix}.
From content providers' perspective, CDNs can deliver their content faster and more efficiently; from local ISPs' perspective, CDNs can reduce the traffic volume from upstream, saving transit costs from their providers. Very often, ISPs do not charge the CDNs for putting servers in their networks. When the video streaming giant Netflix moved online a few years ago, its traffic surged immediately. Now it accounts for up to $32.7\%$ of peak U.S. downstream traffic \cite{CNN_Netflix} and its traffic volume is higher than
that of BitTorrent \cite{BT_Netflix} applications. Netflix used Limelight, one of the biggest CDNs, for content delivery, and later, the Tier-1 Level 3 also obtained a contract to deliver Netflix's traffic. Since most of the Netflix customers are based in the U.S., they often use Comcast, the biggest access ISP, as the last-mile access provider. Interestingly, Comcast managed to enter a so-called paid-peering relationship \cite{clark07} with Level 3 and Limelight, under which the Tier-1 ISP and the CDN have to pay the access ISP for higher bandwidth on the last mile connection.
This has totally {\em reversed} the nominal customer-provider relationship where the Tier-1 ISP was the service provider and should have received payment for connectivity.

It is important to understand how these changes come about, and what the driving factors are behind these changes.
In this work, we model the Internet evolution from a macroscopic view that captures
the cause and effect of the evolution of the individual players in the ecosystem.
Our model expends the traditional view of a single best-effort service model to capture multiple value-added services in the Internet.
The main approaches and contributions are as follows. 
\begin{itemize}
\item We model the preferences and business decisions of the application providers for purchasing Internet services, based on
the application characteristics and the price and quality of the transport services (Section \ref{sec:model}).

\item We characterize the market price and the market share of the Internet transport services by using general equilibrium theory in economics (Section \ref{sec:equilibrium}).

\item We analyze the driving forces of the evolution of the Internet economic ecosystem (Section \ref{sec:dynamics}), which provide qualitative answers (Section \ref{subsec:explanations}) to questions like: {\em 1) Why have the IP transit prices been dropping? 2) Why have the CDNs emerged in the ecosystem? 3) Why has the pricing power shifted to the access ISPs? 4) Why are the large content providers building their own wide-area networks toward users?}

\item We incorporate Internet price and capacity data into our model, and quantitatively fit historical prices and 
    project the future evolution of the ecosystem and its price trends (Section \ref{sec:evolution}).

\item We demonstrate how our model can help the network providers to make business decisions, e.g., capacity expansion and peering decisions, based on the future price projections under various scenarios (Section \ref{sec:evolution_TP_decision}).
\end{itemize}

Our paper sheds new light on the macroscopic evolution of the Internet economic ecosystem and concretely identifies the driving factors of such an evolution. In particular, our model provides a tool to analyze and project the evolutionary trends of the ecosystem. The fundamental understanding of the preferences of application providers and the market equilibrium of the Internet services will also help the business decisions of the application and network transport providers to succeed in this competitive ecosystem.

\section{The Macroscopic AP-TP Model}\label{sec:model}
We start with a macroscopic model of the Internet ecosystem that consists a set of Application Providers (APs) and Transport Providers (TPs). The TPs differ by their service qualities and the prices they charge. We model and analyze the APs' choice of TP based on their own characteristics: how profitable the AP is and how sensitive the AP traffic is to the obtained level of service quality.
In essence, this macroscopic model can help us to understand the decision process of these players in the Internet ecosystem and how these decisions may influence their business relationships.

\subsection{The Application and Transport Providers}
We consider an Internet service market of a geographic region and denote $(\cal M, N)$ as a macroscopic model of the ecosystem, consisting of a set $\cal M$ of TPs and a set $\cal N$ of APs.
The APs provide the content/service for the Internet end-users; the TPs provide the network infrastructure for delivering the APs' data to their end-users.

Our notion of an AP broadly includes content providers, e.g., Netflix, online services, e-commerce, and even cloud services, e.g., Amazon EC2 \cite{amazon}.
Our notion of a TP is based on the APs' point of view. In other words, the transport services provided by the TPs are for the APs to reach their customers/users.
The scope of a TP is {\em broader} than an ISP, and it includes CDNs. 
ISPs, depending on different taxonomies \cite{clark07, dhamdhere08ten, ma11on},
include
1) eyeball/access ISPs that serve the last-mile for end-users,
2) backbone/Tier 1 ISPs that provide transit services for lower tier ISPs,
and 3) content ISPs that serve APs and host content servers.
A TP can be an ISP of any type. Although access and transit ISPs
traditionally do not have business relationships with APs explicitly,
with the emergence of video streaming APs, e.g., Netflix,
we have seen more and more APs' direct or indirect contracts with
the access and transit ISPs. For example, Level 3 contracted with Netflix for
content delivery and Comcast managed to charge Level 3 and Limelight via
paid-peering contracts (for delivering Netflix's traffic
to Comcast's customer base faster) \cite{Norton-playbook}.
Although ``whether ISPs should be allowed to differentiate services/charges
for APs'' is hotly debated under the network neutrality \cite{tim05nn} argument,
legitimate service differentiations will also induce more extensive business
relationships among the APs and ISPs.
In general, a TP can be any facilitator that delivers content to end-users.
An important example of a TP that does not even own network infrastructures in the current Internet ecosystem is
Akamai \cite{Akamai}, which represents the CDNs.

We characterize each TP $I \in \cal M$ by its type, denoted as a triple $(p_I, q_I, \nu_I)$.
$p_I$ denotes the per unit traffic charge for the APs to use TP $I$.
$q_I$ denotes the service quality of TP $I$, e.g., queueing delay or packet
loss probability. Without loss of generality, we assume that $q_I \geq 0$
and smaller values of $q_I$ indicate better quality of services.
$\nu_I$ denotes the bandwidth capacity of TP $I$. 
We characterize each AP $i \in \cal N$ by its utility function $u_i(\cdot)$.
In particular, we define $u_i(p_I, q_I)$ as AP $i$'s utility when it uses
TP $I$, which depends on the service quality $q_I$ and the per unit traffic charge $p_I$.
{\assumption $u_i(\cdot , \cdot)$ is non-increasing in both arguments.\label{assumption:utility}}

{\assumption For any set $\cal M$ of TPs, each AP $i\in \cal N$ chooses to use a TP, denoted as $I_i \in \cal M$,
that satisfies
\[ u_i(p_{I_i}, q_{I_i}) \geq u_i(p_I, q_I), \ \ \forall \ I \in \cal M. \]
\label{assumption:singleton_preference}}
The above assumes that each AP is rational and chooses a TP that provides the
highest utility. Technically, there might exists multiple TPs that
provide the same amount of utility for the AP. We assume that every AP has
certain preference to break the tie and choose one of the TPs.
We further denote ${\cal N}_I \subseteq \cal N$ as the set of APs that
choose to use TP $I$,
or the market share of TP $I$, 
defined as
${\cal N}_I = \{i\in {\cal N} : I_i = I\}$.




Based on Assumption \ref{assumption:utility} and \ref{assumption:singleton_preference}, if two TPs $I,J$ have the same quality, i.e., $q_I=q_J$, then they have to price equally, i.e., $p_I=p_J$; otherwise, the one with higher price will not obtain any market share.
As TPs differ only by price $p_I$ and quality $q_I$ from the APs' perspective,
we aggregate the TPs that have the same value pair $(p_I, q_I)$ into
a single TP with a capacity that equals the summation of individual TPs' capacity.
Similarly, if a TP performs service differentiations, we conceptually
treat it as multiple TPs, each with a service class $(p_I, q_I)$
and the corresponding capacity $\nu_I$.
More precisely, our abstraction of a TP $I$ models a competitive market segment that
provides a quality level $q_I$ and has a total capacity $\nu_I$.

{\lemma Under Assumption \ref{assumption:utility}, if $\cal M$ and $\cal M'$
are identical except for one TP $I$ with $p'_I > p_I$,
then ${\cal N}_I({\cal M'})\subseteq {\cal N}_I({\cal M})$.\\
\label{lemma:AP_monotonicity}}

\smallskip
\noindent {\bf Proof:} Let $i$ be an AP $i\in {\cal N}_I({\cal M'})$. By Assumption \ref{assumption:singleton_preference}, we have $u_i(p'_I,q'_I)\geq u_i(p'_J,q'_J)$ for all $J \in {\cal M'}\backslash \{I\}$.
Since $p'_I > p_I$, we have $u_i(p_I,q_I) = u_i(p_I,q'_I) \geq u_i(p'_I,q'_I) \geq u_i(p'_J,q'_J) = u_i(p_J,q_J)$ for all $J \in {\cal M}\backslash \{I\}$.
This implies that AP $i$ will choose TP $I$ over all the TPs, and therefore, $i\in {\cal N}_I({\cal M})$. This concludes ${\cal N}_I({\cal M'})\subseteq {\cal N}_I({\cal M})$.
\done

Lemma \ref{lemma:AP_monotonicity} implies that when a TP increases (decrease) price unilaterally, fewer (more) APs will choose to
use it.  Intuitively, when TP $I$ increases $p_I$, the utility of each AP in ${\cal N}_I$ does not increase. It is possible that some of them move to other TPs which now provide higher utility tha TP $I$. However, no APs that originally chose other TPs will move to \mbox{TP $I$}.

{\definition[Convexity] The pricing of $\cal M$ is {\em convex} if for
any TPs $I,J,K\in\cal M$ with $q_I < q_K < q_J$,
\[ p_K \leq \eta p_I + (1-\eta) p_J, \]
where $\eta = (q_J-q_K)/(q_J - q_I)$.\\
\label{definition:convexity}}

The above definition is a discrete version of a continuous
convex pricing function. Convex pricing often reflects
the underlying convex cost structure where the
marginal cost monotonically increases with the level of quality.

{\definition[Quasi-Concavity] The utility function $u_i$ is
{\em quasi-concave} if the upper contour sets
$\{(p_i, q_i) \in
\mathbb{R}^2_{+}: u_i(p_i,q_i) \geq u\}$ are
convex for all $u\in\mathbb{R}$.\\
\label{definition:quasi-concavity}}

The quasi-concavity of the utility function implies that if two
choices $(p_1,q_1)$ and $(p_2,q_2)$ provide at least $u$ amount of utility
for AP $i$, then any linear combination of the choices will induce
at least that amount of utility for AP $i$.
In practice, an AP often prefers better quality services until a certain level at which the price becomes a concern.
Combined with a convex pricing, a quasi-concave utility function implies this kind of single-peak preference of the AP as follows.

{\lemma[Single-Peak Preference] When the pricing of $\cal M$ is convex
and $u_i$ is quasi-concave, for any TPs $I,J\in\cal M$
with $u_i(p_I, q_I) > u_i(p_J, q_J)$, then
$u_i(p_J, q_J) \geq u_i(p_K, q_K)$ if $q_I<q_J<q_K$ or $q_I>q_J>q_K$.\\
\label{lemma:one-peak}}

\smallskip
\noindent {\bf Proof:} We first consider the case $q_I<q_J<q_K$. By Definition \ref{definition:convexity}, we know that
$p_J \leq \eta p_I +(1-\eta)p_K$, where $\eta = (q_K - q_J)/(q_K-q_I)$. By Assumption \ref{assumption:utility}, we have
$u_i(p_J, q_J) \geq u_i(\eta p_I +(1-\eta)p_K, q_J) = u_i(\eta p_I +(1-\eta)p_K, \eta q_I +(1-\eta)q_K)$.
By Definition \ref{definition:quasi-concavity}, we have
$u_i(p_J, q_J) \geq u_i(\eta p_I +(1-\eta)p_K, \eta q_I +(1-\eta)q_K) \geq \min (u_i(p_I, q_I),u_i(p_K, q_K))$.
Since $u_i(p_I, q_I) > u_i(p_J, q_J)$, we must have $u_i(p_J, q_J) \geq u_i(p_K, q_K)$.
The derivation of the case $q_I>q_J>q_K$ is similar.
\done

Lemma \ref{lemma:one-peak} gives a condition under which 
if an AP prefers a higher (lower) quality TP $I$ over a lower (higher) quality TP $J$, then it prefers $I$ over any TP whose quality is inferior (superior) to that of $J$. This condition will help us to understand the collective choice of APs of different types in Section \ref{subsection:AP_Choices}.


\subsection{Throughput and Types of the APs}
Although the utility function $u_i$ can be used to model
all the characteristics of AP $i$, the setting does not yet
capture the traffic dynamics and the profitability of the APs.
We model AP $i$'s profitability by
denoting $v_i$ as its per unit traffic revenue.
This revenue is related to the AP's core business,
e.g., online adverting or e-commerce, and we do not
assume how it is generated.
We denote $\lambda_i(\cdot)$ as AP $i$'s throughput function, where $\lambda_i(q_I)$ defines the aggregate throughput of AP $i$ toward its consumers under a quality level $q_I$. Thus, we model any AP $i$'s utility as its total profit
(profit margin multiplied by the total throughput rate), defined by
\begin{equation}
u_i(p_I, q_I) = (v_i-p_I)\lambda_i(q_I).
\end{equation}


{\assumption For any AP $i\in \cal N$, $\lambda_i(\cdot)$ is a non-increasing function with $\alpha_i = \lim_{q_i\rightarrow 0} \lambda_i(q_i)$ and $\lim_{q_i\rightarrow \infty} \lambda_i(q_i)=0$.\label{assumption:throuhgput_rate}}

\smallskip

Assumption \ref{assumption:throuhgput_rate} says that the throughput will not decrease if an AP uses a better service.
$\lambda_i$ reaches a maximum value of $\alpha_i$ when it receives the best quality $q_i = 0$ and decreases to zero if the quality deteriorates infinitely, i.e., $q_i$ tends to $+\infty$.
In particular, we consider the following canonical form of the throughput function: 
\begin{equation}
 \lambda_i(q_I) = \alpha_i e^{-\beta_i q_I},
\end{equation}
where AP $i$'s throughput is characterized by a parameter $\beta_i$ that captures its sensitivity to the received quality $q_I$.

\begin{figure}[ht]
\centering
\includegraphics[width=4in, angle=0]{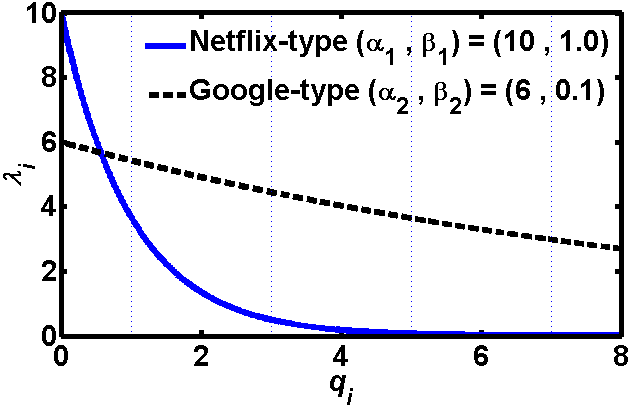}
\caption{Throughput of different type of APs.}
\label{figure:Throughput}
\end{figure}
Figure \ref{figure:Throughput} illustrates
the throughput of two APs with parameters $(\alpha_1, \beta_1) = (10, 1.0)$ and $(\alpha_2, \beta_2) = (6, 0.1)$ under varying service qualities, interpreted as network delays in this case, along the x-axis. 
AP $1$ represents a Netflix-type of application that is more sensitive to delay and has a high maximum rate $\alpha_1 = 10$ Mbps; however, AP $2$ represents a Google-type of query application that is less sensitive to delay. 
We observe that when delay increases, the throughput of delay-sensitive application decreases sharply,
while the delay-insensitive application decreases only mildly.

Because $\alpha_i$ is just a linear scaling factor of the throughput, it does not affect the AP's preference over different TPs.
Consequently, APs with the same $(\beta_i, v_i)$ value pairs will choose the same TP; and therefore, we can conceptually aggregate them as a single AP. Similar to a TP $I$ representing a market segment, each AP $i$ can be interpreted as a group of APs with the same characteristics and $\alpha_i$ represents the aggregate maximum traffic intensity, which depends on the number of APs in the group and the individual traffic intensities.
Although $\alpha_i$ does not play a role in the AP's decision of choosing TPs, we will see later that $\alpha_i$ reflects the demand of the APs and affects the market prices of the TPs.
In summary, based on our throughput model, we define 
\begin{equation}
u_i(p_I, q_I) = (v_i-p_I)\lambda_i(q_I)= \alpha_i(v_i-p_I) e^{-\beta_i q_I}.
\end{equation}
Similar to each TP $I$'s type $(p_I,q_I,\nu_I)$, we can characterize any AP $i$'s type as another triple $(\alpha_i,\beta_i,v_i)$.

\subsection{APs' Choice of Transport Providers}\label{subsection:AP_Choices}
When facing a set $\cal M$ of TPs, each AP $i$'s best choice $I_i$ depends on the price-quality pairs $\{(p_I, q_I):I\in \cal M\}$ and its own characteristics $(\beta_i,v_i)$.

\begin{figure}[ht]
\centering
\includegraphics[width=4in, angle=0]{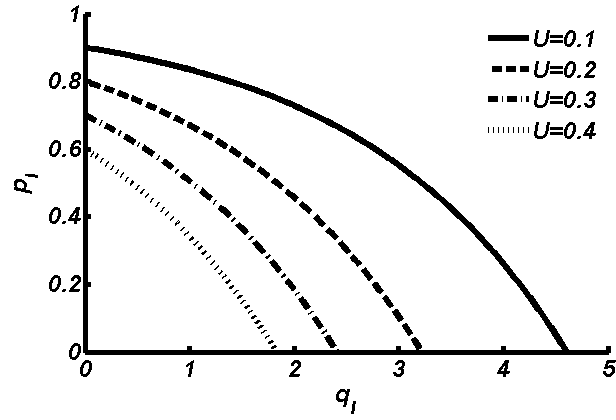}
\caption{Indifference sets of AP $i$, $(v_i,\beta_i)=(1.0,0.5)$.}
\label{figure:upper-contour}
\end{figure}
Given any AP $i$ with $(\beta_i, v_i)$ and a real value $u$, we define the set $\{(p_I, q_I):u_i(p_I, q_I)=u\}$ as the indifference set of TPs that provide $u$ amount of utility for AP $i$. We denote $U$ as the normalized utility defined by $U = u / \alpha_i$ and plot the indifference sets of AP $i$ with $(v_i,\beta_i)=(1.0,0.5)$ in Figure \ref{figure:upper-contour}.
We vary $p_I$ and $q_I$ on the y-axis and the x-axis. Each point $(p_I,q_I)$ on the plane represents a type of TP.
We observe that in order to achieve higher utility, the AP needs a point $(p_I, q_I)$ closer to the origin, which means either the service quality is better, or the charge is cheaper, or both.

\begin{figure}[ht]
\centering
\includegraphics[width=4in, angle=0]{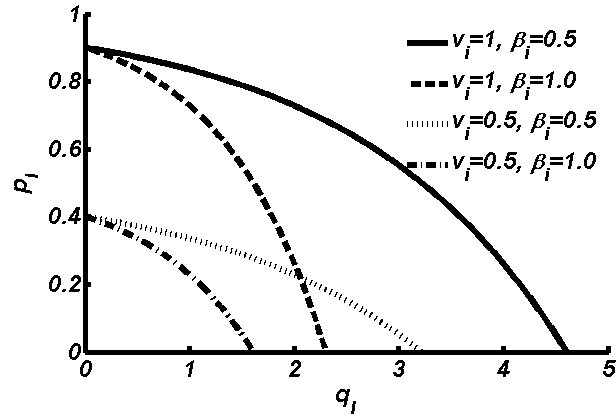}
\caption{Indifference set for $U=0.1$ of different APs.}
\label{figure:upper-contour2}
\end{figure}
In Figure \ref{figure:upper-contour2}, we fixed the normalized utility $U=0.1$ and show the indifference set of different types of APs. We observe that when $\beta_i$ increases, which implies that throughput rate becomes more sensitive to quality, the indifference set shifts from right to left,
showing that the APs require a better service quality to keep its utility. Similarly, when $v_i$ decreases, which implies that
the profitability weakens, the indifference set shifts from top to bottom,
showing that the APs require a lower pricing by TPs in order to keep its utility.

{\theorem For a fixed set $\cal M$ and any two APs $i$ and $j$ with  $\beta_j \geq \beta_i$ and $v_j \geq v_i$, their chosen service qualities satisfy $q_{I_i}\geq q_{I_j}$.
\label{theorem:AP_choice_monotonicity}}

\smallskip
\noindent {\bf Proof:} We prove by contradiction. Assume $q_{I_i} < q_{I_j}$, by Assumption \ref{assumption:utility} and \ref{assumption:singleton_preference}, we know that $p_{I_i} > p_{I_j}$. By Assumption \ref{assumption:singleton_preference}, we further know that
\[ (v_i-p_{I_i})e^{-\beta_i q_{I_i}} \geq (v_i-p_{I_j})e^{-\beta_i q_{I_j}}; \]
\[ (v_j-p_{I_j})e^{-\beta_j q_{I_j}} \geq (v_j-p_{I_i})e^{-\beta_j q_{I_i}}. \]
From the above two inequalities, we can derive
\[ \frac{v_i - p_{I_i}}{v_j - p_{I_i}} e^{-(\beta_i-\beta_j)q_{I_i}} \geq  \frac{v_i - p_{I_j}}{v_j - p_{I_j}} e^{-(\beta_i-\beta_j)q_{I_j}}. \]
However, because $p_{I_i} > p_{I_j}$, we have $ \frac{v_i - p_{I_i}}{v_j - p_{I_i}} < \frac{v_i - p_{I_j}}{v_j - p_{I_j}}$ and because  $q_{I_i} < q_{I_j}$ and $\beta_j\geq \beta_i$, we have $e^{-(\beta_i-\beta_j)q_{I_i}} < e^{-(\beta_i-\beta_j)q_{I_j}}$. By combining both inequalities, we derive the contradiction that 
\[ \frac{v_i - p_{I_i}}{v_j - p_{I_i}} e^{-(\beta_i-\beta_j)q_{I_i}} <  \frac{v_i - p_{I_j}}{v_j - p_{I_j}} e^{-(\beta_i-\beta_j)q_{I_j}}. \]
\done

Theorem \ref{theorem:AP_choice_monotonicity} says that if an AP $j$ is more profitable and more sensitive to service quality than another AP $i$, then the chosen quality of AP $j$ will be at least as good as that of AP $i$.
This property holds regardless how the services are priced.


{\theorem For any $\kappa_1, \kappa_2, \kappa_3 > 0$, and system $(\cal M,N)$, we define a scaled system $({\cal M',N'})$ as
${\cal M'} = \{(\kappa_1 p_I +\kappa_2, q_I / \kappa_3, \nu_I):I\in\cal M\}$ and  ${\cal N'} = \{(\alpha_i, \kappa_3 \beta_i, \kappa_1 v_i+\kappa_2):i\in\cal N\}$, then system $({\cal M',N'})$ satisfies  ${\cal N}_I({\cal M',N'})={\cal N}_I({\cal M,N})$ for all $I\in \cal M$.\label{theorem:scale_invariant}}

\smallskip
\noindent {\bf Proof:} By definition, $u_i(p_I,q_I) = \alpha_i (v_i-p_I)e^{-\beta_i q_I}$. Thus, under the scaled system, we have
\[u_i(p'_I,q'_I) = \alpha'_i (v'_i-p'_I)e^{-\beta'_i q'_I} = \alpha_i (\kappa_1 v_i + \kappa_2 - (\kappa_1 p_I+\kappa_2))e^{-\kappa_3 \beta_i q_I/\kappa_3} =  \alpha_i \kappa_1(v_i-p_I)e^{-\beta_i q_I} = \kappa_1 u_i(p_I,q_I). \]
Since all the utilities of the APs are scaled by $\kappa_1$ in the system $({\cal M',N'})$, their choices of TPs do not change, and as a result, the market share ${\cal N}_I({\cal M',N'})$ of the TPs do not change too.
\done

Theorem \ref{theorem:scale_invariant} says that if 1) the AP profitability $v_i$ and the TP price $p_I$ are linearly scaled in the same way, and/or 2) the quality $q_I$ of the TPs and the sensitivity $\beta_i$ of the APs scale inversely at the same rate, then the APs' choices of TP will not change. This result will help us normalize different systems and make a fair comparison of various solutions.

{\theorem For any $\kappa>0$ and a fixed set $\cal N$ of APs, let ${\cal M'} = \{(p_I, \kappa q_I, \nu_I):I\in\cal M\}$, then for all $i\in\cal N$, 1) $q_{I'_i}\leq \kappa q_{I_i}$ if $\kappa>1$ and 2) $q_{I'_i}\geq \kappa q_{I_i}$ if $\kappa<1$.\label{theorem:scale_variant}}

\smallskip
\noindent {\bf Proof:} We show part 1) by contradiction, and part 2) can be shown by the same arguments.
Assume for some $\kappa>1$, $I'_i = J$ with $q_J > q_{I_i}$. By assumption \ref{assumption:singleton_preference}, we have
\[ (v_i-p_{I_i})e^{-\beta_i q_{I_i}} \geq (v_i - p_J) e^{-\beta_i q_J}; \]
\[ (v_i-p_{I_i})e^{-\beta_i \kappa q_{I_i}} \leq (v_i - p_J) e^{-\beta_i \kappa q_J}. \]
However, the above inequality can be rewritten as 
\[ (v_i-p_{I_i})e^{-\beta_i q_{I_i}} e^{-\beta_i (\kappa-1) q_{I_i}} \leq (v_i - p_J) e^{-\beta_i q_J} e ^{-\beta_i (\kappa-1) q_J}. \]
Because $q_J > q_{I_i}$ and $\kappa>1$, we have $e^{-\beta_i (\kappa-1) q_{I_i}} > e^{-\beta_i (\kappa-1) q_{J}}$. Combined with the condition $(v_i-p_{I_i})e^{-\beta_i q_{I_i}} \geq (v_i - p_J) e^{-\beta_i q_J}$, we have a contradictory condition
\[ (v_i-p_{I_i})e^{-\beta_i \kappa q_{I_i}} > (v_i - p_J) e^{-\beta_i \kappa q_J}. \]
\done

Theorem \ref{theorem:scale_variant} says that if all the qualities in the market deteriorate ($\kappa>1$) linearly at the same rate, APs will not use worse quality TPs than before. The opposite is also true: when qualities improve linearly, 
APs will not use better quality TPs than before.


\begin{figure*}[ht]
\centering
\includegraphics[width=7in, angle=0]{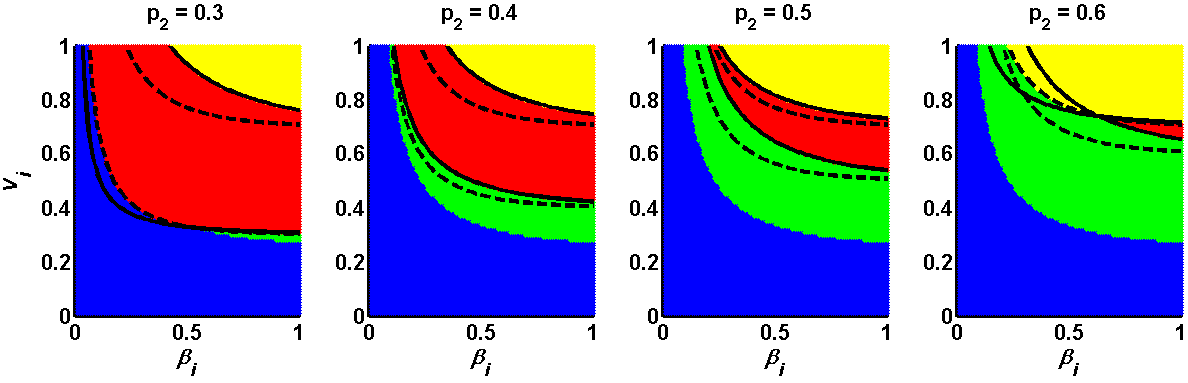}
\caption{Shift of market share for four TPs under $(q_1, q_2, q_3, q_4) = (1,3,5,7)$ and $(p_1, p_3, p_4)=(0.7,0.25,0.1)$.
 }
\label{figure:indifference}
\end{figure*}

\begin{figure*}[ht]
\centering
\includegraphics[width=7in, angle=0]{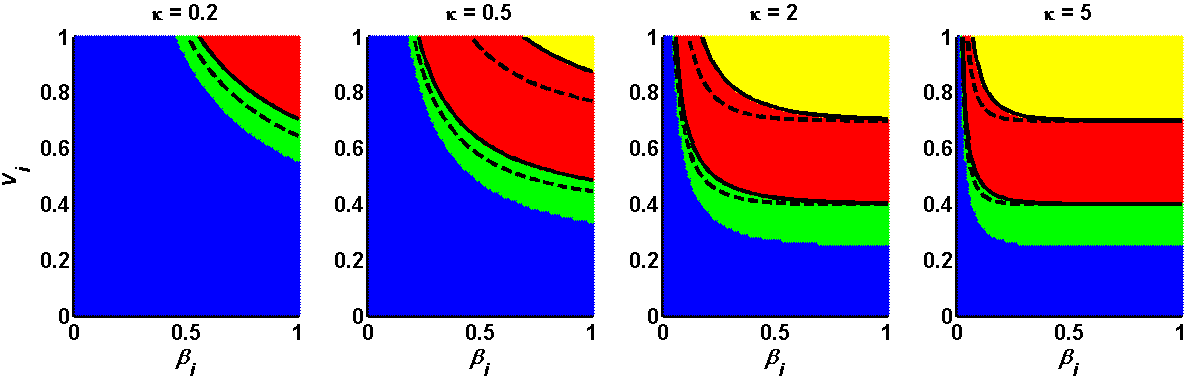}
\caption{Shift of market share for four TPs under $(q_1, q_2, q_3, q_4) = \kappa(1,3,5,7)$ and $(p_1, p_2, p_3, p_4)=(0.7,0.4,0.25,0.1)$.
}
\label{figure:indifference2}
\end{figure*}

With this framework, we can understand the choices made by APs when there
are multiple TPs.  To illustrate, we consider the collective choices of the APs under a market of four TPs. 
In Figure \ref{figure:indifference}, We fix the qualities to be $(q_1, q_2, q_3, q_4) = (1,3,5,7)$ and the prices to be $(p_1, p_3, p_4)=(0.7,0.25,0.1)$ and vary $p_2$ from $0.3$ to $0.6$ in the four subfigures from left to right.
In each subfigure, we vary $\beta_i$  on the x-axis and $v_i$ on the y-axis. Each point $(\beta_i,v_i)$ on the plane represents a type of AP.
The APs located on the top are more profitable and the APs located on the right are more sensitive to the quality of service.
Notice from Figure \ref{figure:Throughput} that a Netflix-type AP $i$, i.e., $\beta_i=1$, would obtain around $40\%$ and $5\%$ of its maximum throughput under quality $q_1$ and $q_2$; however, under $q_3$ and $q_4$, its obtainable throughput almost reaches zero.
Thus, APs with higher value of $\beta_i$ will more
likely choose higher quality TPs. The sets ${\cal N}_1,{\cal N}_2,{\cal N}_3$ and ${\cal N}_4$ are
shown in yellow, red, green and blue respectively.
For example, ${\cal N}_1$ (${\cal N}_4$) represents the set of APs that eventually
choose the TP that provides the highest (lowest) quality with the highest (lowest)
price.
For any $I,J\in \cal M$, we define  ${\cal N}_{IJ} = \{(\beta_i,v_i): u_i(p_I, q_I)=u_i(p_J,q_J)\}$ to be the set of APs that obtain equal utility from $I$ and $J$. In each sub-figure, we plot 
${\cal N}_{12}$ and ${\cal N}_{23}$ in solid lines and ${\cal N}_{13}$ and ${\cal N}_{24}$ in dashed lines.
Thus, Figure \ref{figure:indifference} illustrates
the {\em shift of market shares} for these four TPs when we vary the price $p_2$ of TP 2.

We make the following observations. First, with the increase (decrease) of $p_2$, ${\cal N}_2$ decreases (increases) monotonically (by Lemma \ref{lemma:AP_monotonicity}).
Second, if we keep increasing (decreasing) $p_2$ to $p_1$ ($p_3$), ${\cal N}_2$ (${\cal N}_3$) will become empty (by Lemma \ref{lemma:AP_monotonicity}).
Third, the upper-right APs always choose TPs with better qualities (by Theorem \ref{theorem:AP_choice_monotonicity}).
Finally, when $p_2 = 0.4$ or $0.5$, each set ${\cal N}_i$ forms a distinct band; however, when $p_2 = 0.3$ and $0.6$, we find ${\cal N}_3$ and ${\cal N}_2$ to be isolated regions respectively.
This can be explained by the nature of the pricing of $\cal M$ and the quasiconcavity of the utility function as follows.

{\lemma The utility function $u_i(p_I, q_I) = (v_i-p_I)\lambda_i(q_I)$ is quasiconcave if $\lambda_i(\cdot)$ is in the form of
$\lambda_i(q_I) = \alpha_i e^{-\beta_i q_I}$.\\\label{lemma:quasiconcavity}}

Notice that when $p_2=0.4$, the pricing of $\cal M$ becomes convex and by Lemma \ref{lemma:one-peak} and \ref{lemma:quasiconcavity}, each AP has a single-peak preference among the TPs, where the bands show the preference peaks of the APs.
When $p=0.3$ or $p=0.6$, the non-convexity in pricing induces non-single-peak preferences of some APs. For example, when $p_2=0.3$ ($p_2=0.6$), we can identify APs that prefer TP $2$ and TP $4$ (TP $1$ and TP $3$) over TP $3$ (TP $2$), where ${\cal N}_3$ (${\cal N}_2$) shrinks to be an isolated region.

Let us illustrate the shift of market share when
TPs vary their capacity.
In Figure \ref{figure:indifference2}, we fix the prices
$(p_1, \!p_2, \!p_3, \!p_4)$ $=$
$(0.7,\!0.4,\!0.25,\!0.1)$ and qualities $(q_1,\! q_2,\! q_3,\! q_4)$ $=$
$\kappa (1,\! 3, \!5, \!7)$, and scale the capacities by $\kappa = 0.2, 0.5, 2$ and $5$ from left to right. We observe that when the qualities degrade, APs' choices move to better quality TPs gradually (Theorem \ref{theorem:scale_variant}).

In summary, we presented a framework to help us to analyze (and understand)
the APs' decision on choosing TPs based on each TP $I$'s quality and
price $(q_I, p_I)$, and the AP $i$'s profitability and sensitivity to
quality $(v_i, \beta_i)$.
In reality, the prices of the TPs fluctuate due to competition.
Next, we will study what affect the market prices
and characterize the equilibrium market prices,
which also depends on the traffic intensity
$\alpha_i$ of the APs and the capacity $\nu_I$ of the TPs.

\section{Market Equilibrium} \label{sec:equilibrium}
In this section, we start with the definition of a market equilibrium, under which the prices of the TPs are stable and the claimed service qualities can be achieved when APs choose their best TPs. We then proceed to characterize the market equilibrium and calculate the equilibrium prices.

\subsection{The Existence of Market Equilibrium}
Although any TP $I$ claims to provide service quality $q_I$, it cannot keep its promise if more APs choose this TP than its capacity can support. We model the achieved quality $Q_I(\lambda_I,\nu_I)$ as a function of the actual throughput $\lambda_I$ going through $I$ and its capacity $\nu_I$.
{\assumption The achieved quality $Q_I(\lambda_I,\nu_I)$  for any TP $I\in\cal M$ is non-decreasing in $\lambda_I$ and non-increasing in $\nu_I$. \label{assumption:quality}}
{\definition A set ${\cal X}\subseteq \cal N$ of APs is {\em feasible} for TP $I$ with quality $q_I$, if 
$Q_I \big(\lambda_I({\cal X}),\nu_I\big) \leq q_I$,
where $\lambda_I({\cal X}) = \sum_{i\in{\cal X}}\lambda_i(q_I)$ defines the induced throughput of the set ${\cal X}$ of APs under quality $q_I$.
\label{definition:feasible_set}}

In a market $\cal M$ of TPs, each TP would adjust its strategies to accommodate its customer APs' traffic demand and keep its service quality promise. For example, if the current capacity of TP $I$ cannot support quality $q_I$, it might 1) expend its capacity $\nu_I$, 2) increase price $p_I$, or 3) reduce the quality level $q_I$. Next, we define a market equilibrium where the APs' demand are {\em feasible} and the TPs' strategies are {\em stable}.

{\definition Let $p^{min}_I$ be the cost (or minimum price) of TP $I$. Let $\cal M'$ be identical to $\cal M$ except for $p'_I \neq p_I$ for some $I\in \cal M$ and ${\cal N}^{'}_I$ be the set of APs choosing TP $I$ under $\cal M'$.
A system $(\cal M,N)$ forms a {\em market equilibrium} if 1) all APs' aggregate demands are feasible, i.e.,
\[Q_I \big(\lambda_I({\cal N}_I),\nu_I\big) \leq q_I, \ \ \forall I\in\cal M, \]
and 2) each price $p_I$ maximizes the utilization of capacity for acceptable throughput at TP $I$, i.e.,
for any $p'_I\geq p^{min}_I$ with the corresponding ${\cal N}^{'}_I$ satisfying $Q_I \big(\lambda_I({\cal N}^{'}_I),\nu_I\big) \leq q_I$,
\[\lambda_I({\cal N}^{'}_I)\leq \lambda_I({\cal N}_I).\] \label{definition:equilibrium}}

\vspace{-0.1in}
One way to understand the above definition of a market equilibrium is that given a set $\cal N$ of APs and a set
\mbox{$\{q_I\!:\!I\!\in\!\cal M\}$} of service qualities for them to choose from, the price $p_I$ and capacity $\nu_I$ of each market segment should be consistent in that 1) when the APs make their choices of TP, their expected service quality can be achieved and, 2) the capacities of the TPs are not under-utilized, unless the charge $p_I$ reaches the TP's cost $p^{min}_I$. If APs' quality expectations are not fulfilled, their choices of TP will change. Furthermore, if capacity $\nu_I$ is under-utilized with $p_I\!>\!p^{min}_I$, then the market segment $I$ is not correctly priced.
That being said, we assume that none of the market segment is controlled by a monopoly, which might want to under-utilize capacity and keep a higher price for profit-maximization. We will summarize and discuss the limitations of our model in Section \ref{sec:limiation}. The interesting aspect here is that although $p_I$, like all other prices, mainly depends on the supply $\nu_I$ and the demand ${\cal N}_I$ of the APs, all the TPs (or market segments) are correlated, which serve substitutions for the APs.

In practice, the TPs might not have enough capacities to accommodate all
APs. As a result, market prices will rise and some APs cannot afford
the prices and will not use any of the TPs. However, under
Assumption \ref{assumption:singleton_preference}, each AP needs
to choose a TP even it cannot afford to use any of the TPs,
so a market equilibrium might not exist under this assumption.
To fix this minor technical issue,
we make the following assumption to allow any AP not to use any of the TPs if they all induce
negative utilities. 
{\assumption There always exists a dummy TP $D\in\cal M$ with quality $q_{D}=\infty$ and price $p_{D}=0$.
\label{assumption:dummy_ISP}}

\smallskip

By Assumption \ref{assumption:throuhgput_rate}, quality $q_{D}$ always induces zero throughput for any AP, and therefore, the dummy TP guarantees a zero utility and can accommodate as many APs as possible in equilibrium. Effectively, the set ${\cal N}_D$  models the APs that cannot afford to use any TP in the market in reality.

{\theorem For any fixed set $\cal N$ of APs and any set $\cal M$ of TPs with fixed values of $p^{min}_I, q_I$ and $\nu_I$ for all $I\in\cal M$, there exists a set $\{p_I:I\in\cal M\}$ of prices that makes $\cal (M,N)$ a market equilibrium.
\label{theorem:equilibrium_existence}}

\smallskip
\noindent {\bf Proof:} The proof of existence of a market equilibrium is a constructive one. We can start with $p_I=p^{min}_I$ for all $I\in\cal M$, and for each overloaded TP, we increase its price until its capacity is fully utilized. For each such a step, when overloaded TP $I$'s price is increased, APs will move to other TPs, making them possibly overloaded too. Thus, the prices of the TPs will be monotonically non-decreasing during the process, and therefore, will converge to a market equilibrium.
\done


Although TPs might be able to adopt new technologies to improve or differentiate their services, the quality that they can provide is often physically constrained by the nature of the TP, for example, if a TP is a Tier 1 ISP, it cannot guarantee end-to-end delays for the customers unless the access ISP's link is not congested.
Similarly, although TPs might execute a long-term capacity planning, the supply of capacity does not change in a small time scale.
Compared to service quality and capacity, market prices change more frequently and easily. Theorem \ref{theorem:equilibrium_existence} says that even in a small time-scale where prices adapt to market conditions, prices might still converge to an equilibrium, which reflects the short-term market structure of the Internet ecosystem.

\subsection{Characteristics of a Market Equilibrium}
In theory, one might find multiple sets of prices that make $(\cal M,N)$ a market equilibrium. For example, from any existing equilibrium, one might find a TP $I$ such that with only a small change in $p_I$, no APs will change their choices. This new price also constitutes a market equilibrium.
In practice, these price differences can happen by two reasons. First, even without a monopoly in a market segment, oligopolistic providers might implicitly collude on the price so that they keep a relatively high price simultaneously. When one of them starts to reduce price, the price of that segment will converge to a lower price. Second, the preferences of the APs are quite different so that the price change in one segment might not affect the demand choices of the APs.

{\definition A market equilibrium $(\cal M,N)$ is {\em competitive} if there does not exist any $p^{min}_I\leq p'_I<p_I$ with the corresponding ${\cal N}^{'}_I$ satisfying $Q_I \big(\lambda_I({\cal N}^{'}_I),\nu_I\big) \leq q_I$. \label{assumption:competitive_equilibrium}}

\smallskip

If the AP types are very diverse or each market segment consists of many competing providers, one can focus on the above definition of a competitive market equilibrium. Technically, a competitive market equilibrium might not exist, since the minimum price might not exist when all the feasible equilibrium prices form an open set. However, prices in practice have a minimum unit, e.g., one cent, and we can always find such a competitive market equilibrium.

We will later show how to calculate competitive market equilibria.
We would like to point out that our model is not limited to competitive market
equilibria, i.e., if a segment $I$ is not competitive enough, we can use a higher price for $p_I$.
As a result, competitive equilibrium prices might be biased downward if
the real market structure is not perfectly competitive;
nevertheless, our qualitative results do not depend on whether the
market equilibrium is competitive or not.

{\theorem Let ${\cal N'} = \{(\kappa\alpha_i,\beta_i, v_i):i\in\cal N\}$ and ${\cal M'} = \{(p_I, q_I, \kappa \nu_I):I\in\cal M\}$ for some $\kappa>0$.
If $(\cal M,N)$ is a market equilibrium and the quality function $Q_I(\cdot,\cdot)$s are homogenous of degree $0$, i.e.,
$Q_I(\lambda_I, \nu_I)=Q_I(\kappa \lambda_I, \kappa \nu_I)$, $\forall \kappa>0, I\in\cal M$,
then $({\cal M'}, \cal N')$ is a market equilibrium too.
\label{theorem:linear_scale}}

\smallskip
\noindent {\bf Proof:} Since $\alpha_i$ is a linear scaling factor, it does not affect the choices of the APs and the market shares of the TPs. As a result, when $\alpha_i$s are scaled by $\kappa$, the aggregate throughput $\lambda_I$s are scaled by $\kappa$ too. Because $Q_I(\cdot,\cdot)$s are homogenous of degree $0$, when $\nu_I$s are scaled by $\kappa$ as the same time, the achieved quality values do not change in the scaled system. By the monotonicity of $Q_I$ of Assumption \ref{assumption:quality}, we know that the equilibrium conditions of Definition \ref{definition:equilibrium} do not change, and therefore, $({\cal M'}, \cal N')$ is a market equilibrium too.
\done

\smallskip
Theorem \ref{theorem:linear_scale} says that if the quality only depends on the ratio of the incoming traffic rate and the capacity, then when the number of APs (and their traffic intensity) and the capacities scale at the same speed, the original market equilibrium prices will remain in equilibrium. If we consider the queueing delay as the quality metric, because of statistical multiplexing, the average queueing delay reduces when both arrival rate and service rate scales up at the same rate. In this case, Theorem \ref{theorem:linear_scale} also implies that each TP $I$ can accept more and more traffic for a fixed delay $q_I$, and as a consequence, the market prices will move downward in a new equilibrium.

\subsection{Calculating Market Equilibrium Prices}
We denote $\mu_I$ as the maximum throughput that TP $I$ can accept when it can still fulfill the quality $q_I$, defined as
\begin{equation}
\mu_I = \arg \max_{\lambda_I} \ Q_I(\lambda_I,\nu_I) \leq q_I.
\end{equation}

For instance, if the quality metric is the average queueing delay under M/G/1 systems and TP $I$ implements a FIFO scheduling policy, by the Pollaczek-Khinchine mean formula,
\[ Q_I\big(\lambda_I,\nu_I\big) = \frac{\lambda_I}{\nu_I-\lambda_I} E[R], \]
where $E[R]$ is a constant that denotes the expected residual service time of jobs. If we want $\lambda_I$ to be feasible, we need
\[\frac{\lambda_I}{\nu_I-\lambda_I} E[R] \leq q_I \Rightarrow \lambda_I\leq \frac{q_I}{E[R]+q_I}\nu_I=\mu_I. \]
We define $\eta_I = \mu_I/\nu_I$ as the maximum acceptable throughput per unit capacity, or the conversion factor from raw capacity to achievable throughput.
Notice that given a fixed capacity $\nu_I$, the smaller delay TP $I$ wants to provide, the smaller maximum amount of traffic it can accept.
For the M/G/1 case, $\eta_I$ tends to $0$ when the required quality $q_I$ tends to $0$, which also shows a convex cost structure for the TP.

Based on the monotonicity of $Q_I$ (Assumption \ref{assumption:quality}), a market equilibrium can be characterized by using $\mu_I$s as follows.

{\definition A system $(\cal M,N)$ forms a {\em market equilibrium} if for all TP $I$, 1) $\lambda_I({\cal N}_I) \leq \mu_I$, and 2) there does not exist $p'_I \geq p^{min}_I$ with the corresponding ${\cal N}^{'}_I$ satisfying $\lambda_I({\cal N}_I) < \lambda_I({\cal N}^{'}_I) \leq \mu_I$. \label{definition:alternative}}


\smallskip

Based on the above alternative definition of a market equilibrium, we can calculate the competitive equilibrium prices without evaluating $Q_I$ repeatedly as follows.
\begin{tabbing}
\centering
\hskip 0.05in \=xx\=x\=x\=x\=x\kill
\rule{5.3in}{0.5mm}\\
{\bf Calculate Price Equilibrium$\big(\cal N$, $\{p^{min}_I, q_I, \nu_I:I\in\cal M\}\big)$} \\

1. \> \> \tbltxt{{\it Set } $p_I = \infty$ {\it for all TP } $I\in\cal M$;} \\

2. \> \> \tbltxt{{\it Calculate } $\mu_I$ {\it for all TP } $I\in\cal M$ based on $q_I$ and $Q_I$;} \\

3. \> \> \tbltxt{{{\bf while} {\it there exists } $p'_I \in [p^{min}_I, p_I)$ {\it such that} $\lambda_I({\cal N}_I)\leq \lambda_I({\cal N}^{'}_I)\leq \mu_I$ } }\\

4. \> \> \> \> \tbltxt{{\bf set} $p_I = p'_I$; }\\

5. \> \> \tbltxt{{\bf return} $\{p_I:I\in\cal M\}$;}\\
\rule{5.3in}{0.5mm}
\end{tabbing}

In the above algorithm, we do not restrict which TP $I$ to choose in step 3 if multiple TPs satisfy the condition. However, any sequence of updates will make the price vector converge, because each price component $p_I$ will only be decreasing monotonically until convergence.
Similarly, we can also set $p_I=p^{min}_I$ for all TPs, and the price vector will increase monotonically until convergence.

Based on Theorem \ref{theorem:scale_invariant} and \ref{theorem:linear_scale}, we also have the following result.


{\corollary Let ${\cal N'} = \{(\kappa\alpha_i, \kappa_3\beta_i, \kappa_1 v_i + \kappa_2):i\in\cal N\}$ and ${\cal M'} = \{(\kappa_1 p_I+\kappa_2, q_I/\kappa_3, \nu'_I):I\in\cal M\}$  for positive $\kappa, \kappa_1, \kappa_2, \kappa_3$ with $\mu'_I=\kappa \mu_I$ for all $I\in\cal M$.
If $(\cal M,N)$ is a market equilibrium, then $({\cal M'}, \cal N')$ is a market equilibrium.\label{corollary:linear_scale}}

\smallskip
\noindent {\bf Proof:} By Theorem \ref{theorem:scale_invariant}, we know that the choices of APs and the market shares of the TPs will not change in both systems. By the same arguments for proving Theorem \ref{theorem:linear_scale}, we know that when 
$\alpha_i$s are scaled by $\kappa$ and the effective capacity $\mu_I$s are scaled by $\kappa$ at the same time, the actual quality $Q_i$s do not change. As a result, the equilibrium conditions of Definition \ref{definition:equilibrium} do not change.
\done

\smallskip

Although the prices of the TPs influence the APs' choices, which further affect the capacity utilization of the TPs, equilibrium prices are the fixed points in which both the APs' choices and the TPs' prices do not change.
However, external factors could move the resulting equilibrium. 
In the next section, we will study these fundamental driving forces for the evolution of the Internet economic ecosystem. By understanding these factors, we will know why the market prices change and why certain evolutions happen. 

\section{Price Dynamics in Equilibrium}\label{sec:dynamics}
In this section, we look deeper into the qualitative dynamics of the equilibrium market prices.
In particular, we explore how the different characteristics of the APs and
the TPs can affect the market prices in equilibrium.

\subsection{Evaluation Setting}
Each AP $i$ is characterized by three parameters $(\alpha_i,\beta_i,v_i)$; each TP $I$ is characterized by three parameters $(p_I, q_I, \nu_I)$. To make a fair comparison between equilibrium prices under different settings, we carefully normalize the system parameters as follows.
We define $v_{max} = \max \{v_i:i\in\cal N\}$, $\beta_{max} = \max \{\beta_i:i\in\cal N\}$, and $p_{min} = \min \{p^{min}_I:I\in\cal M\}$.
Based on Theorem \ref{theorem:scale_invariant}, we normalize any system $(\cal M,N)$ by factors $\kappa_1 = 1/(v_{max}-p_{min})$, $\kappa_2 = p_{min}/(v_{max}-p_{min})$, and $\kappa_3 = 1/\beta_{max}$. As a result, we normalize each $\beta_i$ or $v_i$ within the interval $[0,1]$ and the equilibrium prices will also be scaled accordingly with $[0,1]$. If $p^{scaled}_I$ is the derived market equilibrium price in the normalized system,
we can recover the real market price $p_I$ as
\[p_I = (v_{max}-p_{min})p^{scaled}_I+p_{min}.\]
When the normalized price $p^{scaled}_I$ tends to $0$, it reflects that the real market price $p_I$ goes down to the cost $p_{min}$; when $p^{scaled}_I$ tends to $1$, it reflects that the real market price $p_I$ goes to the maximum AP profitability $v_{max}$.
We describe the TPs' capacity in terms of the maximum acceptable rates $\mu_I$s.
We define $\alpha = \sum_{i\in \cal N}\alpha_i$, $\mu = \sum_{I\in\cal M} \mu_I$ and the ratio $\rho = \mu / \alpha$. Based on Corollary \ref{corollary:linear_scale}, any price equilibrium sustains when $\alpha_i$s and $\mu_I$s scale at the same rate. Thus, we normalize the APs' aggregate maximum traffic intensity $\alpha$ to be $1$. We define $\sigma_I = \mu_I/\mu$ as the capacity share of TP $I$, and under the normalized system, each TP $I$ has $\mu_I = \sigma_I \rho$.

After the above normalization, we can describe any system by the following four parameters:
\begin{enumerate}
\item a set of qualities $\{q_I:I\in\cal M\}$,
\item the normalized aggregate capacity $\rho$,
\item the distribution of $\alpha_i$ over the domain $[0,1]^2$ of $(\beta_i,v_i)$,
\item the capacity distribution  $\{\sigma_I:I\in\cal M\}$.
\end{enumerate}

\begin{figure}[ht]
\centering
\includegraphics[width=4.5in, angle=0]{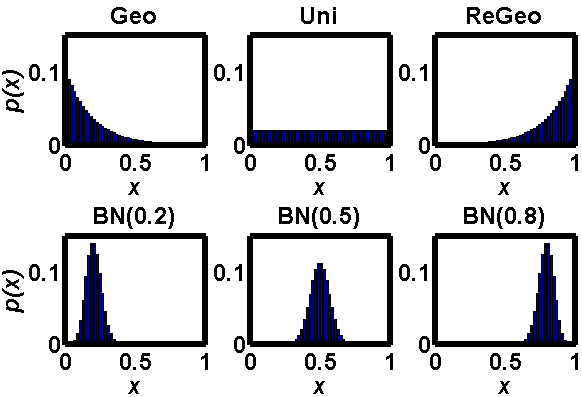}
\caption{Common distributions: geometric, uniform, reversed geometric, binomial with $p=0.2, 0.5$ and $0.8$.}
\label{figure:common_distributions}
\end{figure}

We focus on three different quality types:
1) $q_A$, the highest quality for real-time content delivery,
2) $q_B$, medium quality, mostly for web applications,
and
3) $q_C$, the best-effort quality, mostly for elastic traffic.
As analyzed in \cite{tiers11}, IP transit markets will be quite efficient if two tiers of services are provided; thus,
$q_B$ and $q_C$ can be considered as the higher and lower tier services of such an IP transit market.
To differentiate the three qualities, we set
$q_A:q_B:q_C=1:5:25$. We vary $\rho$ from $0$ to $1$, where the system's total capacity varies from extremely scarce to abundant.
We discretize the AP domain with $50$ levels of $v_i$ and
$\beta_i$, which forms $2500$ types of APs.
We assume that APs' profitability and quality-sensitivity
follow probability distributions $F_v$ and $F_\beta$ respectively,
and $\alpha_i$ follows the joint distribution of $F_v$ and $F_\beta$.
We use the various distributions in Figure \ref{figure:common_distributions} for $F_v$ and $F_\beta$.
For instance, when a geometric distribution $Geo$ is used to describe $F_\beta$, it models
the scenario where most of the AP traffic are elastic and the amount of quality-sensitive traffic decreases exponentially with its sensitivity level $\beta_i$. The binomial distributions $BN(p)$ are often used to approximate a normal distribution of the profitability $v_i$, or quality sensitivity $\beta_i$, where $p$ determines the mean value.


\subsection{Impact of TP Capacity on Prices}
In this subsection, we study how the capacities of the TPs affect the equilibrium prices.
We initially set $(q_A,q_B,q_C) = (0.2, 1, 5)$.
We will evaluate how the quality may impact
the equilibrium prices in the next subsection.

\begin{figure}[!ht]
\centering
\includegraphics[width=6.3in, angle=0]{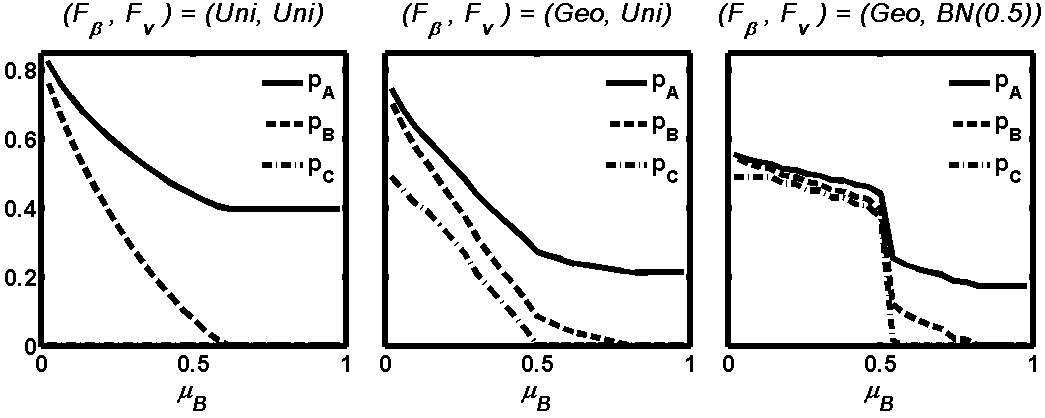}
\caption{Shift in market prices as $\mu_B$ varies: with $(q_A, q_B, q_C)=(0.2, 1, 5)$, $\mu_A=0.05$ and $\mu_C=0.25$.}
\label{figure:capacity_effect1}
\end{figure}

\begin{figure}[!ht]
\centering
\includegraphics[width=6.3in, angle=0]{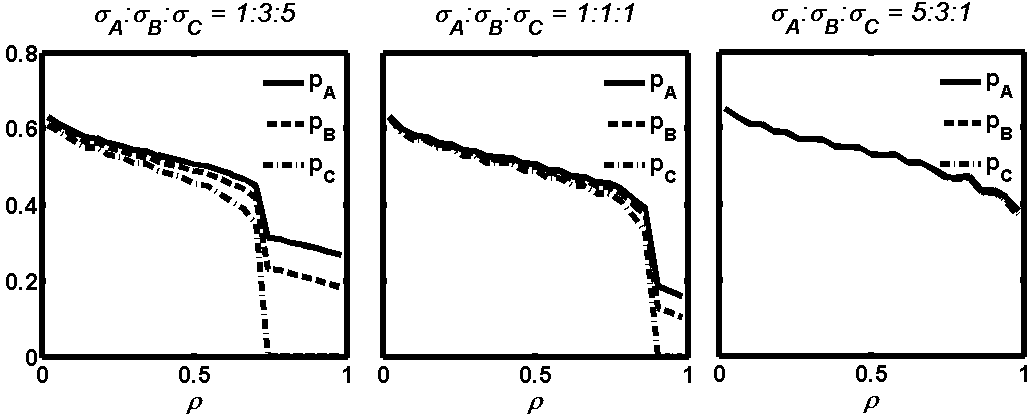}
\caption{Shift in market prices as $\rho$ varies: with $(q_A, q_B, q_C)=(0.2, 1, 5)$ and $(F_\beta, F_v) = (Geo, Uni)$.}
\label{figure:capacity_effect2}
\end{figure}

In Figure \ref{figure:capacity_effect1}, we fix $\mu_A=0.05, \mu_C=0.25$ and vary $\mu_B$ from $0$ to $1$ along the x-axis. The three sub-figures show the equilibrium prices when $\alpha_i$ follows the joint distributions of $(F_\beta, F_v) = (Uni, Uni), (Geo, Uni)$ and $(Geo, BN(0.5))$ respectively.
We observe that when $\mu_B$ is scarce, equilibrium price $p_B$ is close to (but
strictly less than) the price $p_A$ of its upper class TP.
When $\mu_B$ increases, $p_B$ diverges from $p_A$ and moves to the price $p_C$ of its lower class TP. When $\mu_B$ becomes abundant, its market price goes down to the minimum price after $p_C$. In general, when the capacity of a particular TP, i.e., $\mu_B$, increases, it drives all equilibrium prices down; however, the prices of higher quality TPs, e.g., $p_A$, might not go down to the minimum price. 

In the rest of this section, we often use $F_\beta = Geo$, which models the case where more APs were elastic, and $F_v = BN(0.5)$, which approximates that the AP profitability follows a normal distribution centered at $v_i=0.5$. Note that our qualitative results do not depend on these settings.

In Figure \ref{figure:capacity_effect2}, we vary the system capacity $\rho$ from $0$ to $1$ along the x-axis. $\alpha_i$ follows the joint distribution $(F_\beta, F_v) = (Geo, BN(0.5))$. The sub-figures show the equilibrium prices when the capacity ratio
\mbox{$\sigma_A:\sigma_B:\sigma_C$} equals \mbox{$1:3:5$}, \mbox{$1:1:1$} and \mbox{$5:3:1$} respectively. In all
three cases, when the total capacity $\rho$ is small, all equilibrium prices are very close and high.
When we increase $\rho$, all market prices drop.
By comparing the price curves across the three subfigures, we observe that when the capacity share of the higher class TP is smaller (the left subfigure), 1) the three market prices differ more from each other, 2)  $p_C$ drops faster, and 3) all the prices drop to the minimum price faster than the other two cases. Because 
price differences exist in practice, we will use $\sigma_A:\sigma_B:\sigma_C = 1:3:5$ in the rest of this section.

\medskip
\noindent{\bf Lessons (the TP capacity effects on prices) leaned:}
\begin{itemize}
\item Capacity expansion drives market prices down. 
\item The capacity expansion of a particular TP $I$ would affect not only its own price $p_I$, but also other TPs' prices, due to the substitution effect of TP $I$ to other TPs.
\item When TP $I$'s capacity share $\sigma_I$ is small (big), its market price $p_I$ is close to the price of its next higher (lower) class TP.
\end{itemize}

\subsection{Impact of TP Quality on Prices}
Let us explore how the quality $q_I$ of the TPs may
affect the equilibrium prices.
We use the setting that the capacity distribution
follows $\sigma_A:\sigma_B:\sigma_C = 1:3:5$
and $\alpha_i$ follows the joint distribution
$(F_\beta, F_v) = (Geo, BN(0.5))$.

\begin{figure}[!ht]
\centering
\includegraphics[width=6.3in, angle=0]{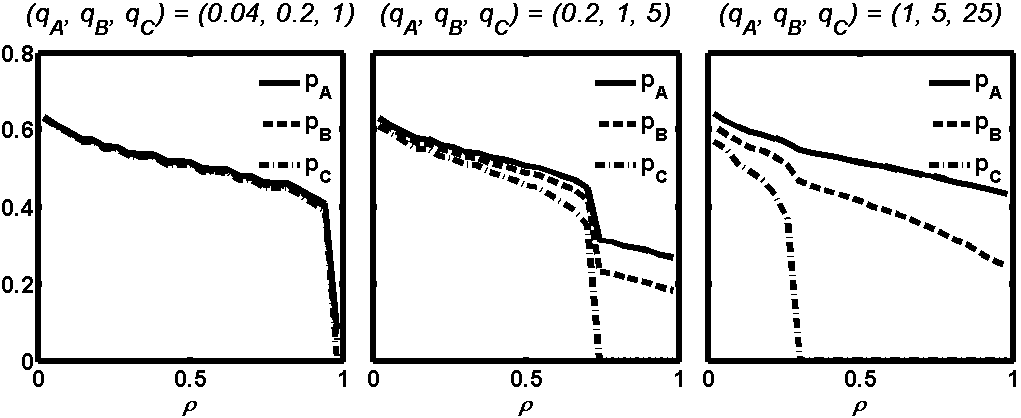}
\caption{Shift in market prices as $\rho$ varies: with $(q_A, q_B, q_C)=\kappa (0.2, 1, 5)$ where $\kappa = 0.2, 1$ and $5$.}
\label{figure:quality_effect1}
\end{figure}

In Figure \ref{figure:quality_effect1}, we keep the quality ratio  $q_A : q_B = q_B : q_C = 1: 5$ and use $(q_A, q_B, q_C) = \kappa(0.2, 1, 5)$, where $\kappa$ equals $0.2, 1$ and $5$ in the three subfigures. We vary the system capacity $\rho$ from $0$ to $1$ along the x-axis.
\begin{figure}[!ht]
\centering
\includegraphics[width=6.3in, angle=0]{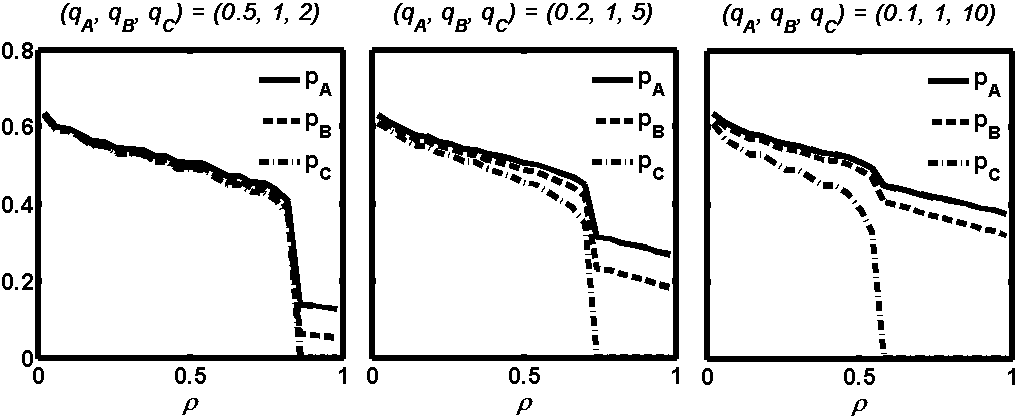}
\caption{Shift in market prices as $\rho$ varies: with $q_B=1$, $q_A\!:\! q_B\! =\! q_B\!:\! q_C\! =\! 1\!:\!\kappa$ where $\kappa \!=\! 2,\! 5,\!100$.}
\label{figure:quality_effect2}
\end{figure}
We observe that when all the TPs improve their quality by the same ratio, i.e., $\kappa=0.2$, the market prices of the TPs are very close; when all the TPs degrade their quality by the same ratio, i.e., $\kappa=5$, the market prices of the TPs diverge greatly. This observation can be explained by Theorem \ref{theorem:scale_variant}. When $\kappa$ decreases and all qualities are improved, more APs will choose
lower class TPs, which move the prices of the lower class TPs upward and the prices of upper class TPs downward. As a result, all TPs prices will move closer. On the other hand, when $\kappa$ increases and all qualities are degraded, more APs will choose to upper class TPs, which move the prices of the upper class TPs upward and the prices of lower class TPs downward. This will further diverge the price differences among the TPs with different qualities.

In Figure \ref{figure:quality_effect2}, we keep $q_B = 1$ and vary the quality ratio $q_A:q_B = q_B:q_C = 1:\kappa$, where $\kappa$ equals $2,5$ and $10$.
We observe that the price differences are positively correlated with the quality ratio. In particular, when quality ratio is high, e.g., $\kappa=10$, the price of the lowest class TP, i.e., $p_C$, drops earlier and sharper when the total capacity $\rho$ expands. At the same time, higher class TPs can still maintain a non-zero market price even after $p_C$ drops down the minimum price. The general trend is that when the quality ratio keeps increasing, the price curves will move higher and toward the left.
In the rest of this section, we will often use the quality ratio $1:5$ and $(q_A, q_B, q_C) = (0.2, 1, 5)$ for our evaluations. Again, our qualitative results do not depend on this setting.

\medskip
\noindent {\bf Lessons (the TP quality effects on prices) leaned:}
\begin{itemize}

\item The market prices of the TPs would be close to (far from) one another if the quality ratio is small (big) or/and the overall qualities of the market are high (low).

\item In reality, the qualities provided by the TPs are becoming better and better, which implies that market prices for different services might converge.
\item High-end market segments can still maintain a price difference if they can differentiate their quality from the lower class TPs substantially.
\end{itemize}
Next, we will see that the TP price differences also depend on the demand side: the characteristics of the APs.

\subsection{Impact of AP Wealth on Prices}
Let us explore how the profitability distribution $F_v$ 
may affect the equilibrium prices.
We still keep $\sigma_A:\sigma_B:\sigma_C=1:3:5$ and $(q_A,q_B,q_C) = (0.2, 1, 5)$.
$\alpha_i$ follows the joint distribution of $F_\beta$ and $F_v$, where $F_\beta$ is distributed as $Geo$.

\begin{figure}[!ht]
\centering
\includegraphics[width=6.3in, angle=0]{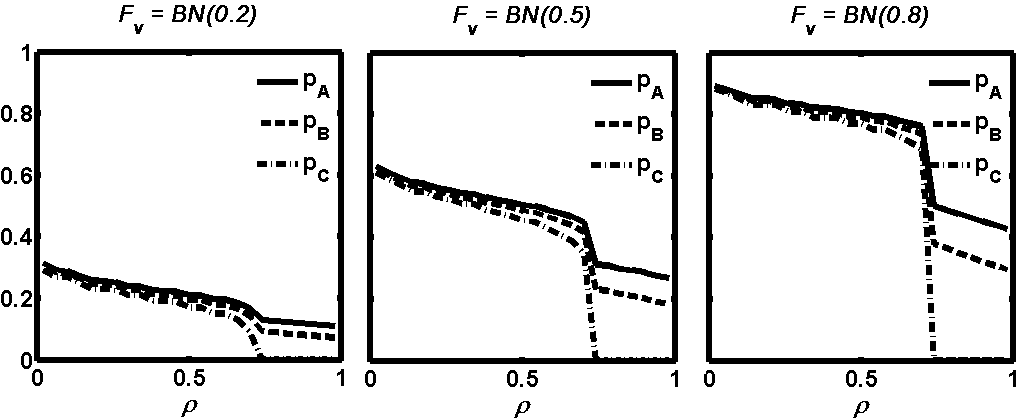}
\caption{Shift in market prices as $\rho$ varies: $(q_A,q_B,q_C)\! =\! (0.2,\! 1,\! 5)$, $\sigma_A\!:\!\sigma_B\!:\!\sigma_C\!=\!1\!:\!3\!:\!5$, $F_\beta \!=\! Geo$.
}
\label{figure:wealth_effect1}
\end{figure}

\begin{figure}[!ht]
\centering
\includegraphics[width=6.3in, angle=0]{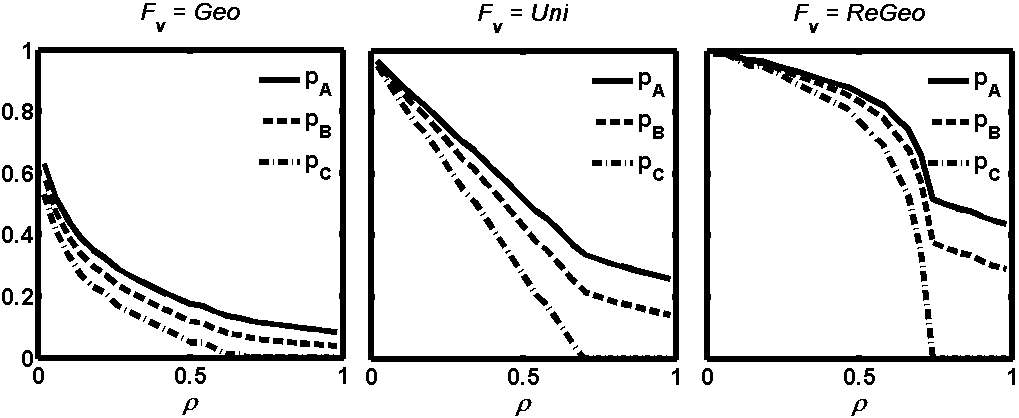}
\caption{Shift in market prices as $\rho$ varies: $(q_A,q_B,q_C)\! =\!(0.2,\! 1,\! 5)$, $\sigma_A\!:\!\sigma_B\!:\!\sigma_C\!=\!1\!:\!3\!:\!5$, $F_\beta \!=\! Geo$.
}
\label{figure:wealth_effect2}
\end{figure}

In Figure \ref{figure:wealth_effect1}, we vary $\rho$ from $0$ to $1$ along the x-axis and plot the equilibrium prices where the profitability distribution $F_v$ follows a binomial distribution $BN(p)$ parameterized by $p=0.2,0.5$ and $0.8$ respectively.
By doing this, we simulate the normal distributions of the APs' wealth varying the mean value from small to large. We observe that despite the difference in mean profit of the APs, $p_C$ drops to the minimum price at the same time.
The price curves in all cases keep the same shape; however, they scale differently on the vertical axis.
This indicates that the market prices depend on how much the APs are able to pay for the services, and how they demand for the TPs based on their values of $(\beta_i,v_i)$.

In Figure \ref{figure:wealth_effect2}, we vary $F_v$ to be $Geo$, $Uni$ and $ReGeo$.
We observe that the shapes of the price curves are very different: prices decrease convexly, linearly and concavely in the three subfigures. In general, how fast the prices drop depends on the density of the APs whose profitability are around that price range, and the shape of the curves look like the complimentary cumulative distribution function (CCDF) of $F_v$.

\medskip
\noindent {\bf Lessons (the AP wealth effects on prices) leaned:}
\begin{itemize}
\item The market prices of the TPs are positively correlated with the mean profitability of the APs.
\item At a certain price range where the density of the APs is high (low), more (less) competition among the APs drives the prices close to (far below) their profitability.
\end{itemize}

\subsection{Impact of AP Quality-Sensitivity on Prices}

In this subsection, we study how the quality sensitivity distribution $F_\beta$ affects the equilibrium prices. We set $\sigma_A:\sigma_B:\sigma_C=1:3:5$ and $\rho=0.5$. In the following cases, $F_\beta$ follows a binomial distribution $BN(p)$, where we vary the parameter $p$ along the x-axis. By doing this, we simulate the cases where the APs become more and more sensitive to quality when the mean sensitivity increases with $p$.

\begin{figure}[!ht]
\centering
\includegraphics[width=6.3in, angle=0]{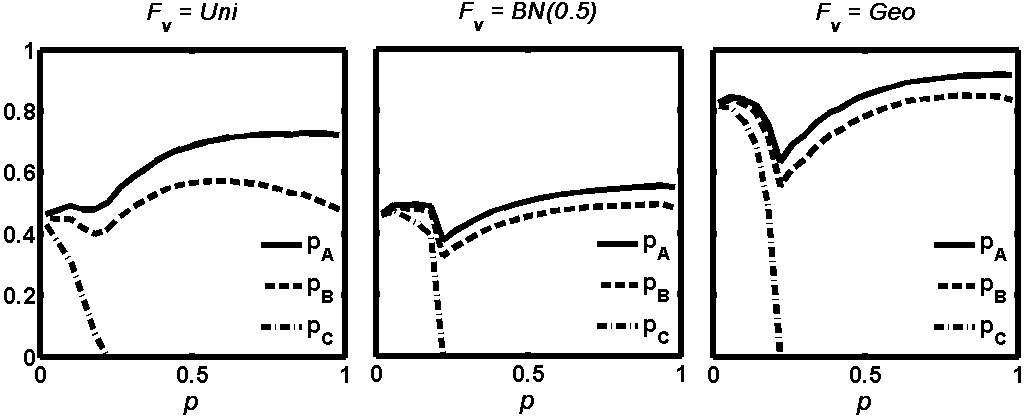}
\caption{Shift of market prices when we vary AP's sensitivity to quality: with
$\sigma_A:\sigma_B:\sigma_C=1:3:5$, $\rho=0.5$, $(q_A,q_B,q_C) = (0.2,1,5)$ and $F_\beta=BN(p)$.
}
\label{figure:sensitivity_effect1}
\end{figure}

\begin{figure}[!ht]
\centering
\includegraphics[width=6.3in, angle=0]{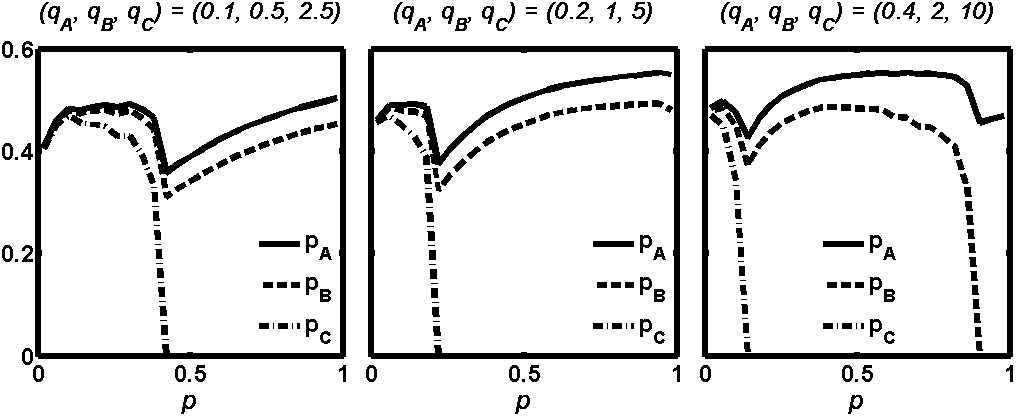}
\caption{Shift of market prices when we vary AP's sensitivity to quality: with
$\sigma_A:\sigma_B:\sigma_C=1:3:5$, $\rho=0.5$, $F_v=BN(0.5)$ and $F_\beta=BN(p)$.
}
\label{figure:sensitivity_effect2}
\end{figure}

In Figure \ref{figure:sensitivity_effect1}, we fix $(q_A,q_B,q_C) = (0.2,1,5)$ and vary $F_v$ to be $Uni$, $BN(0.5)$ and $Geo$ in the three sub-figures. We observe that although the profitability distribution affect the absolute price values, the shape of the price curves look similar. When the quality
sensitivity of the APs increases, the lowest quality service price, i.e., $p_C$, drops sharply and quickly. Although $p_A$ and $p_B$ drops accordingly with $p_C$, after $p_C$ reaches the minimum price,
both $p_A$ and $p_B$ rebound. With further increase in quality sensitivity, $p_B$ shows a trend to decrease slowly; however, $p_A$ always stays at a high level. When the APs become more sensitive to quality, more and more APs start to move to higher class TPs. As a result, the capacity $\mu_C$ becomes under-utilized, which also drives $p_C$ down very quickly. Although $p_C$'s drop pulls down the overall market prices, more APs move to higher class TPs, which make TPs $A$ and $B$ in demand, and therefore, keep $p_A$ and $p_B$ steadier. After $p_C$ reaches the minimum price, $p_C$ stops decreasing.
As the APs' quality sensitivity keeps increasing,
even the minimum market price of $p_C$ becomes relatively expensive to the APs. This makes even more APs move to TP $A$ and $B$ and drives $p_A$ and $p_B$ upward. 

In Figure \ref{figure:sensitivity_effect2}, we fix $F_v = BN(0.5)$ and vary the qualities to be $(q_A,q_B,q_C) = \kappa (0.2, 1, 5)$, where $\kappa=0.5, 1$ and $2$. We observe the same trends as in Figure \ref{figure:sensitivity_effect1} that $p_C$ drops quickly and sharply to the minimum price as the APs' quality sensitivity increases. As $\kappa$ increases, all the price curves move to the left and the price drop of lower class TPs becomes quicker and sharper. This also coincides with the observations made in Figure \ref{figure:quality_effect1} that when the qualities degrade, the price of the lower class TP drops much quicker.

In the above illustrations, we vary the distribution $F_\beta$. It is also possible that all the APs' sensitivity increase by 
$\beta'_i=\xi \beta_i$ for some $\xi>1$. By Theorem \ref{theorem:scale_invariant}, we can rescale the system by $\kappa_3 = 1/ \xi$, as if the APs keep their quality sensitivity constant and all the qualities become poorer. By Theorem \ref{theorem:scale_variant} and the TP quality effect result in Figure \ref{figure:quality_effect1}, we also conclude that more APs will prefer higher quality TPs and the price of the lower quality TPs will drop sharply.

\medskip
\noindent {\bf Lessons (the AP quality-sensitivity effects on prices) leaned:}
\begin{itemize}
\item When the APs become more sensitive to the service quality, the price of lower class TPs will drop quickly.

\item When the price of the lowest quality TP goes down to its cost, the prices of higher quality TPs might increase due to their relatively cheap prices and high demand.

\end{itemize}

\subsection{Internet Evolution: Some Explanations}\label{subsec:explanations}
By understanding the factors that drive the market equilibrium, we reason about the evolution of the Internet ecosystem and reach plausible answers to the questions raised in Section \ref{section:intro}.
We do not claim that our answers below are exhaustive and the limitations of our model will be discussed in Section \ref{sec:limiation}.

\noindent {\bf 1) Why have the IP transit prices been dropping? }
The capacity effect tells that the price drop can be a consequence of the capacity expansion of the transit providers. Compared to the capacities at the last-miles, the capacity in the backbone grows faster than demand and is abundant \cite{weber-pricing}. Also, the price drop in better quality services, i.e., CDN prices, will drive the transit prices further down. The quality effect tells that when the transit quality differs a lot from the CDN services, the prices will diverge greatly. The wealth effect tells that since the majority of the elastic APs might not be very profitable, transit providers cannot fully utilize its capacity and charge a high price at the same time. This is also why they are looking for providing value-added and differentiated services. Last, the AP quality-sensitivity effect tells that when AP traffic becomes more and more inelastic, e.g., the surge of Netflix traffic, lower quality service will become less valuable and therefore its price will drop quickly.

\noindent {\bf 2) Why have the CDNs emerged in the ecosystem? } The capacity effect tells that when the capacity of higher quality service is small, it can maintain a price difference with the lower quality services. The quality effect tells that if a CDN service's quality differs a lot from the transit services, it can be priced much higher. When the capacity of the transit market was limited and priced high, the demand for even higher quality service drove the price for potential CDN services even higher. This explains why CDNs emerged in the first place.
The wealth effect tells that when the APs' profitability is not high, the market prices cannot be high. However, due to the low cost structure of the CDNs, they can still help small APs who could not afford the infrastructure to support large demand.
The AP quality-sensitivity effect further tells that with the traffic being more and more sensitive to quality, the price of high quality CDN can sustain at a high level.

\noindent {\bf 3) Why has the pricing power shifted to the access ISPs? } This can be partially explained by the AP quality-sensitivity  effect and the TP quality effect. When the AP traffic becomes more and more sensitive to service quality, they are more willing to pay for the higher quality services. Because access ISPs are physically closer to the users, their service quality is naturally much better than other providers who have to go through the access ISPs to reach the end-users anyways. Consequently, the difference in service quality makes it possible for the access ISPs to charge services at higher prices. Furthermore, Comcast's monopolistic position in the U.S. market could be another reason, under which its price will be set higher than the competitive market price under Definition \ref{assumption:competitive_equilibrium}.

\noindent {\bf 4) Why are the large content providers building their own wide-area networks toward users?} Mostly because the APs become more sensitive to service quality, they cannot rely on the transit providers to deliver content. 
As high quality services are limited and access ISPs would obtain more pricing power, large APs might consider establishing their own networks toward users as a cheaper alternative than paying access ISPs for better services in the future.

\section{Internet's Economic Evolution}\label{sec:evolution}

Besides understanding how each isolated factor might affect the market prices, we incorporate
ground truth data \cite{PeeringDB, craig10internet,globalInternetGeography, CDNPricing}, e.g., the historical trends of the TPs' capacity expansion and the APs' characteristics, and project {\em possible future price dynamics} of the Internet ecosystem.
Through this, our model can help the TPs make various long-term business decisions.
Let us demonstrate this.


We take a macroscopic view and categorize network services as two types: ${\cal M}=\{A,B\}$. 
$B$ models the IP transit service that provides interconnection based on ``best-effort''; while $A$ models the CDN or private peering type of service that provides better service quality than $B$.
We categorize the APs as three types:  ${\cal N}=\{a,b,c\}$. 
$a$ models the video or realtime interactive applications that are very sensitive to quality. $b$ models the web applications that are elastic but more tolerate to quality than type $a$ applications.  $c$ models the inelastic applications, e.g., email and P2P file download. 

By Theorem \ref{theorem:scale_invariant} and Corollary \ref{corollary:linear_scale}, we know that when quality and the sensitivity parameters scale inversely, the equilibrium remains the same; therefore, without loss of generality, we set $q_B=1$ as the baseline best-effort quality level. We set the quality sensitivity parameters to be $(\beta_a, \beta_b, \beta_c) = (10, 1, 0.1)$.
Under this setting, type $a$ APs would only obtain $e^{-10} \approx 4.5 ^{-5}$ of their maximum throughput under $q_B$, which implies that the best-effort service cannot support quality sensitive applications. Also, under $q_B$, a type $b$ AP could get $e^{-1} \approx 37\%$ of its maximum throughput; however, a type $c$ AP could get $e^{-0.1} \approx 90\%$ of its maximum throughput.
When measured by delay, the quality of service (QoS) for realtime applications often require the delay to be at the order of milliseconds \cite{Xiao-QoS}, compared to the best-effort service delays at the order of seconds. Thus, we choose $q_A = 0.01$ to reflect the same order of magnitude of service difference. As a result, even type $a$ APs would obtain $e^{-0.1} \approx 90\%$ of their maximum throughput under the better quality level $q_A$.

Next, we try to estimate the capacity of the TPs on the Internet. We take the Equinix Internet Exchange at New York (Equinix-NY) as a reference market and estimate the capacities based on the data provided by PeeringDB \cite{PeeringDB}.
At the end of year $2011$, there were $102$ ISPs listed on at Equinix New York Exchange in PeeringDB, among which $44$ use {\em Open} peering policy and the remaining $58$ use either {\em Selective} or {\em Restricted} peering policy.
The total capacity was around $21$ Tbps, among which the ISPs using Open peering policy contributed $7$ Tbps and the remaining ISPs contributed $14$ Tbps.
Since {\em Selective} and {\em Restricted} policies are used for private and often paid-peering agreements, we set $\nu_A$ and $\nu_B$ to be $14$ and $7$ Tbps, for the reference time of the year $2011$.

From the Global Internet Geography \cite{globalInternetGeography} report, between $2007$ to $2011$, the international Internet capacity increased six-fold and the bandwidth to the U.S. had increased nearly $50$ percent per year. To a first approximation, we assume that the capacity $\nu$ of the TPs increases $50\%$ per year.
We define $\alpha = \alpha_a+\alpha_b+\alpha_c$ and $\omega_a$, $\omega_b$ and $\omega_c$ as the weight of the throughput upper bound of each application type. Given $\alpha$ and the weight of AP $i$, we obtain $\alpha_i$ as
\[ \alpha_i = \frac{\omega_i}{\omega_a+\omega_b+\omega_c}\alpha, \quad \forall \ i=a,b,c.\]
Based on the observed traffic distribution of various applications in \cite{craig10internet}, we set $(\omega_a,\omega_b,\omega_c) = (2\%, 75\%, 23\%)$ for the year $2007$, and assume that the weight for video ($\omega_a$), web ($\omega_b$) and inelastic applications ($\omega_c$) increase at an annual growth rate of $150\%$, $50\%$ and $20\%$ respectively.
Notice that IP transit prices are often quoted for per Mbps-month, while CDN prices are often quoted for per terabit.
If capacity is fully utilized $24/7$, \$$1$ per Mbps-month can be translated into \$$0.386$ per terabit. 
We assume that the maximum per unit traffic revenue for the APs is \$$10$ Mbps-month and the APs' revenue are uniformly distributed.

\subsection{A First Approximation Benchmark}

We use our macroscopic model to fit the historical prices starting from $2007$ and project future Internet prices. In a first approximation, we choose the following parameters.
\begin{enumerate}
\item $\alpha$ at year $2007$ (denoted as $\alpha_{07}$) equals $10$ Tbps. 
\item $\alpha$ increases at an annual growth rate $r_{\alpha} = 22\%$.
\item $\eta_A = \mu_A/\nu_A = 0.3$ and $\eta_B = \mu_B/\nu_B = 0.9$.
\end{enumerate}

\begin{figure}[!ht]
\centering
\includegraphics[width=3.0in, angle=0]{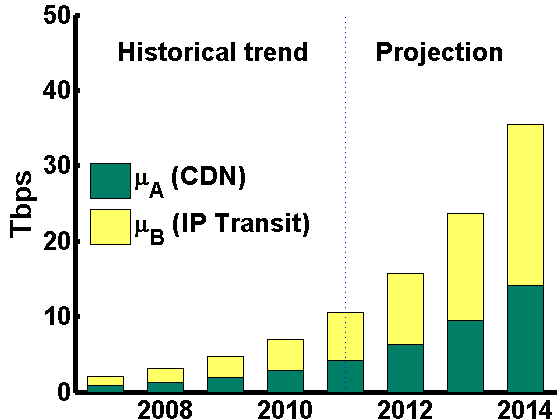}
\includegraphics[width=3.0in, angle=0]{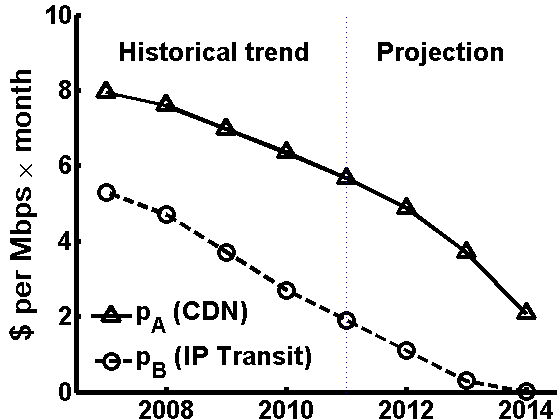}
\includegraphics[width=3.0in, angle=0]{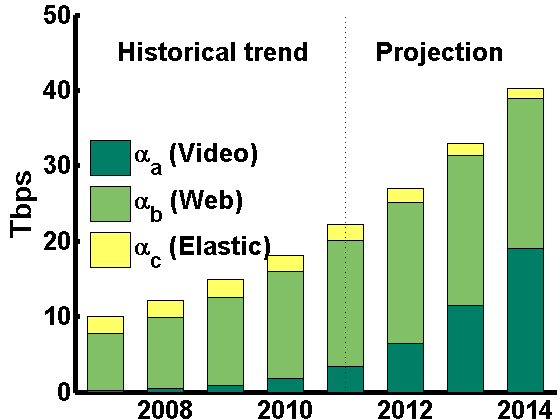}
\includegraphics[width=3.0in, angle=0]{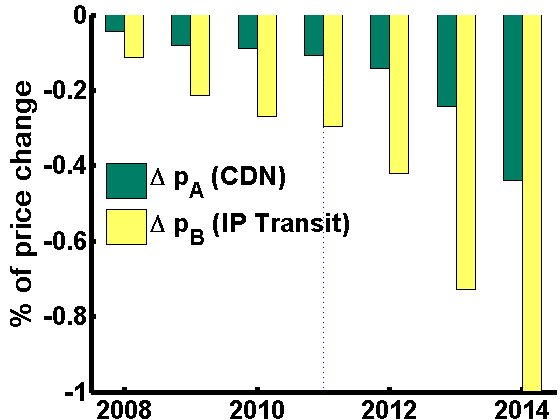}
\caption{Historical price and future price projection.}
\label{figure:benchmark}
\end{figure}

In Figure \ref{figure:benchmark}, the upper left subfigure plots the achievable throughput for the CDN ($\mu_A$) and IP transit ($\mu_B$) services from $2007$ to $2014$ and the lower left subfigure plots the maximum demand $\alpha_a$, $\alpha_b$ and $\alpha_c$ for the same time period.
The upper right subfigure plots the price dynamics of both IP transit and CDN services and the lower right subfigure plots the percentage of price change for both services.
We observe that average price drop from $2007$ to $2011$ is approximately $20\%$, which coincides with the price drop surveyed in the Global Internet Geography \cite{globalInternetGeography} report. Also, the price of IP transit is below \$$2$ per Mbps-month, very close to the mean of IP transit prices, where the lowest price fell to \$$1$ per Mbps-month.

Compared to the video delivery pricing \cite{CDNPricing}, our price projection shows that the CDN price drops around $8\%$ annually from $2007$ to $2011$, and reaches \$$5.67$ per Mbps-month, or \$$2.18$ per terabit. This price is lower than the \$$7.5$ per terabit price for APs with volume of $5$PB data and the price drop is slower than the observed $20\%$ price drop in the CDN industry \cite{CDNPricing}. The difference could come for two reasons: 1) since CDN service charges based on traffic volume, we cannot assume that the APs would always use the capacity $24/7$, and therefore, the CDN providers should charge some premium on top of the basic per Mbps per month charge, 2) in contrast to our competitive model for CDN service, the industry might be less competitive and could charge a much higher price; therefore, when the industry becomes more competitive, we expect to see much sharper price drops.

Based on the trend from $2007$ to $2011$, our model projects that both the IP transit and CDN prices will further drop, at an even faster rate, and IP transit price will drop to its minimum price. Of course, this projection is based on the assumption that the capacity of the TPs will keep expanding at the $50\%$ annual rate. We will further discuss potential trends of future prices in a later subsection.

\subsection{Sensitivity of the Benchmark}
In this subsection, we show the sensitivity of our price projection with respect to the chosen parameters.

\begin{figure}[!ht]
\centering
\includegraphics[width=5.28in, angle=0]{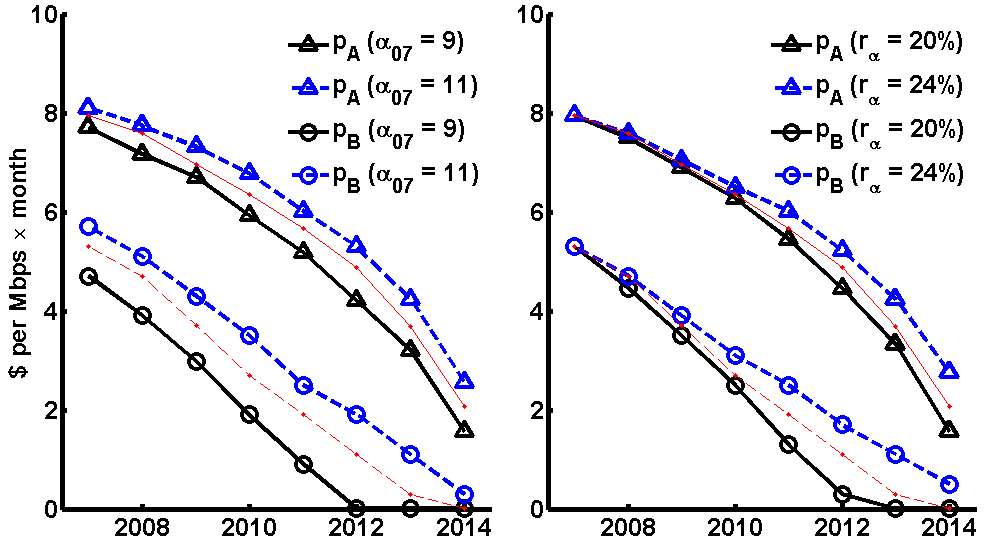}
\caption{Sensitivity to initial demand $\alpha_{07}$ and rate $r_{\alpha}$.}
\label{figure:projection_alpha}
\end{figure}

First, we want to see how the demand parameter $\alpha$ affects the price dynamics.
Figure \ref{figure:projection_alpha} shows a projection of service prices when the initial value $\alpha_{07}$ and the growth rate $r_{\alpha}$ change. In the left subfigure, we vary $\alpha_{07}$ to be $9$ and $11$ Tbps compared to the benchmark value of $10$ Tbps. We observe that the prices are positively correlated with $\alpha_{07}$. In the right subfigure, we vary the growth rate $r_{\alpha}$ to be $20\%$ and $24\%$ compared to the benchmark value of $22\%$. We observe that the prices again are positively correlated with $r_{\alpha}$. Both tells that when the demand increases, so do the prices. 

\begin{figure}[!ht]
\centering
\includegraphics[width=5.28in, angle=0]{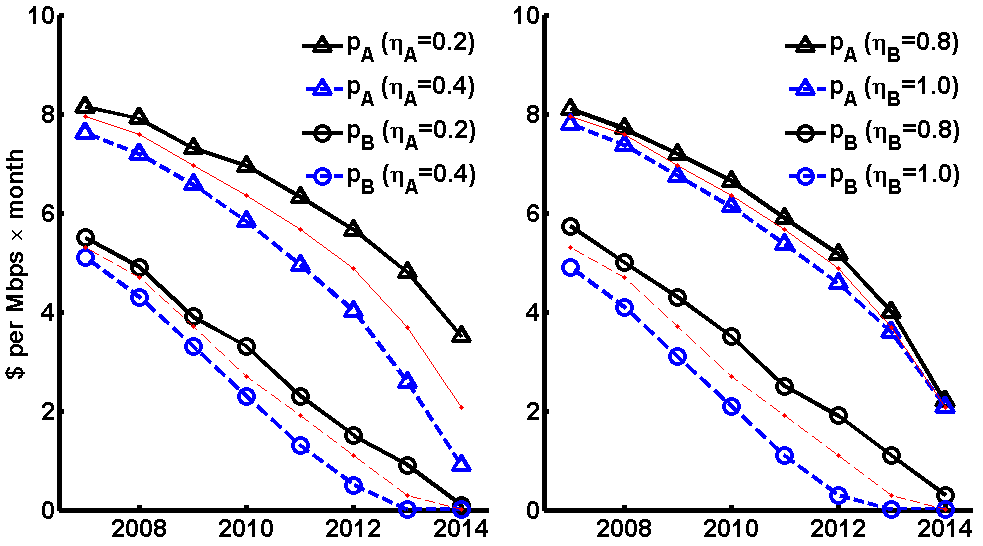}
\caption{Sensitivity to capacity utilization $\eta_A$ and $\eta_B$.}
\label{figure:projection_mu}
\end{figure}

Second, we want to see how the capacity utilization factor $\eta_I = \mu_I/\nu_I$ affects the price dynamics.
Figure \ref{figure:projection_mu} shows a projection of Internet service prices when $\eta_A$ and $\eta_B$ vary from the benchmark.
In the left subfigure, we vary $\eta_A$ to be $0.2$ and $0.4$ compared to the benchmark value $0.3$.
In the right subfigure, we vary $\eta_B$ to be $0.8$ and $1.0$ compared to the benchmark value $0.9$.
We observe that the all prices are negatively correlated with the capacity utilization factors. Also, the IP transit prices are sensitive to both $\eta_A$ and $\eta_B$; while the CDN prices are only sensitive to its utilization factor $\eta_A$.



\subsection{Price Projection and TP Business Decisions}
\label{sec:evolution_TP_decision}

Now, we demonstrate that by using the price projection from our model, we can help the TPs to make business decisions on 1) how aggressive the TPs should expend their capacity, and 2) whether the TPs should/would tend towards {\em Open} or {\em Selective} peering policies. 

\begin{figure}[!ht]
\centering
\includegraphics[width=5.28in, angle=0]{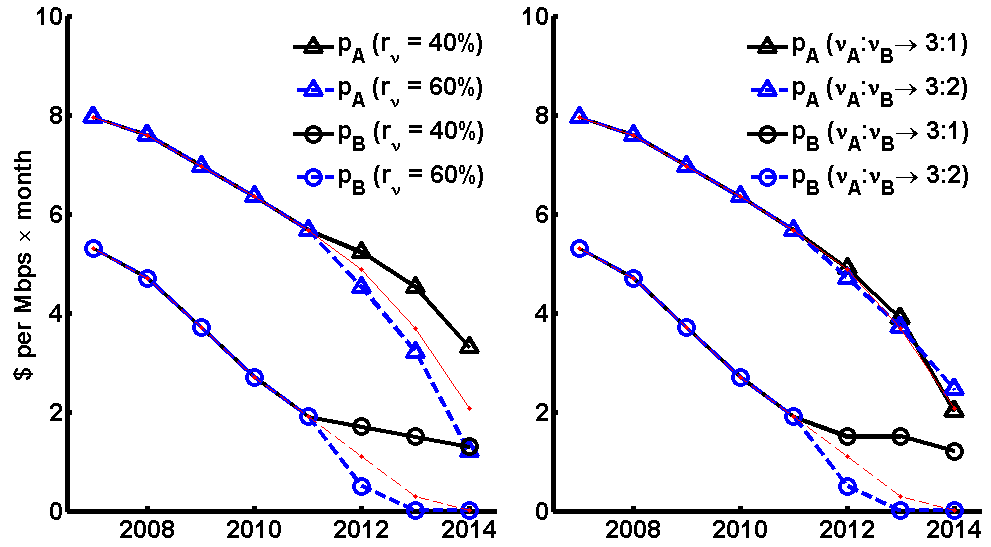}
\caption{Price projection under various capacity ratios $\nu_A:\nu_B$ and capacity expansion rate $r_{\nu}$.}
\label{figure:projection_future}
\end{figure}

To answer the first question, we vary the capacity growth rate of the TPs $r_{\nu}$ to be $40\%$ and $60\%$, compared to the historical benchmark rate $50\%$ and plot the price projections in the left subfigure of Figure \ref{figure:projection_future}.
We observe that when the capacity grows at $60\%$ per year, both the CDN and IP transit price drop fast and the IP transit price will down to its cost next year; however, when the growth rate is $40\%$, the IP transit price will be decreasing at a very slow rate. These observations tell us that the ISPs providing IP transit services might want to slow down their investment in capacity expansion; however, CDN providers and ISPs that sell private-peering and QoS might want to continue to expand their capacity when their profit margins are still above zero. As the price of IP transit drops, we believe that the investment in the transit capacity will slow down, which will also stabilize the price of the IP transit services.

To answer the second question, we vary the capacity ratio $\nu_A:\nu_B$ from the benchmark ratio $2:1$ ($14$ Tpbs $:7$ Tbps) to $3:1$ and $3:2$ for the year $2014$ in the right subfigure. These two projections model the scenarios where ISPs will tend to be more {\em Selective} and more {\em Open} in their peering policies respectively.
We observe that if more ISPs are going to use an {\em Open} peering policy, the IP transit price will drop to its cost quickly; otherwise, the IP transit price will get closer to the CDN price and be stable. This observation implies that ISPs would have strong incentives to move towards {\em Selective} peering policies if possible, which coincides with the reality that the access ISP, Comcast, started to use private peering exclusively.


In summary, we predict that although the CDN price will still be dropping in the coming years, the price of IP transit will be more stable. Furthermore, the capacity expansion will slow down and more ISPs will tend to use {\em Selective} rather than {\em Open} peering policies in the near future.


\subsection{Limitations of the Model}\label{sec:limiation}
Although we have demonstrated potential usages of our model,
we want to mention the limitations of the model so as to avoid misinterpreting the results obtained from our macroscopic model.

First of all, our general equilibrium model implicitly assumes that each market segment is competitive. In practice, some market segment could be lack of competition and form a monopolistic or oligopolistic market structure.
Thus, the real market prices will be higher than what our model predicts.
Second, our equilibrium model does not capture the off-equilibrium and transit dynamics that could happen in practice.
Third, our model is in nature macroscopic, and it does not capture detailed information like peering agreement, topology, traffic patterns and etc. Nevertheless, our model does capture the type of different services the TPs provide via implicitly encoding all the relevant information into the quality level $q_I$. From the APs' point of view, they do care about {\em quality} rather than other details of the TPs.
Fourth, since our focus is on the transit/CDN market, our model does not intend to capture the end-user market aspects. For example, modeling the bundle of access services and other service differentiations are out of scope.
Last but not the least, our macroscopic model provides some qualitative reasons for the Internet evolution, which we do not claim to be exhaustive. There might be additional factors/reasons that are not captured by our model, e.g., the lack of competition in the market.


\section{Related Work}\label{sec:related}
Many empirical studies have been tracking the evolution of the Internet using measurements and public data sets \cite{craig10internet,dhamdhere08ten,flattening08,broadband11,empirical03}. Labovitz et al. \cite{craig10internet} measured the inter-domain traffic between 2007 and 2009, and observed the changes in traffic patterns as well as the consolidation and disintermediation of the Internet core. Gill et al. \cite{flattening08} collected and analyzed traceroute measurements and showed that large content providers are deploying their own wide-area networks. Dhamdhere et al. \cite{dhamdhere08ten} confirmed the consolidation of the core of the Internet, that brings the content closer to users. Akella et al. \cite{empirical03} used measurements to identify and characterize non-access bottleneck links in terms of their location, latency and available capacity. At the edge of the Internet, Sundaresan et al. \cite{broadband11} studied the network access link performance measured directly from home gateway devices. We focus on a macroscopic model of the Internet ecosystem that captures the application traffic going through the network transport service providers.

Many works \cite{peerornot,clark07,ma11on,Motiwala12cost,tiers11,lodhi12GENESIS,Haddadi11Modeling} focused on the modeling perspective of the Internet evolution. Chang et al. \cite{peerornot} presents an evolutionary model for the AS topologies. Lodhi et al. \cite{lodhi12GENESIS} used an agent-based model to study the network formation of the Internet. Motiwala et al. \cite{Motiwala12cost} used a cost model to study the Internet traffic. Valancius et al. combined models and data to study the pricing \cite{tiers11} structure of the IP transit market. Faratin et al. \cite{clark07} and Ma et al. \cite{ma11on} studied the evolution of the ISP settlements.
In this work, we take a holistic view and analyze the business decisions and evolutions of the APs and TPs altogether.

\section{Conclusions}\label{sec:conclusions}
In this paper, we proposed a network aware, macroscopic model to explain the evolution of the Internet.
This model captures
1) the business decisions of the APs,
2) the pricing and competition of the TPs, and
3) the resulting market equilibrium of the ecosystem.
By analyzing how the AP characteristics (i.e., traffic intensity, profitability and sensitivity to service quality),
and the TP characteristics (i.e., quality, price and capacity, affect the market equilibrium), we obtain fundamental understanding of
why historical and recent evolutions of the Internet have happened.
With further estimations of the trends in traffic demand, capacity growth and quality improvements, our model can also project the future evolution of the Internet ecosystem. This model provides a tool for the Internet players to better understand their business and risks, and help them to deal with their business decisions in the complicated and evolving ecosystem.

{
\bibliographystyle{abbrv}
\bibliography{paper}
}

\end{document}